\documentclass[twocolumn,longauthor]{aastex61}
\usepackage{apjfonts}

\received{September 13, 2017}
\accepted{January 15, 2018}

\shorttitle{Dense Regions in Supersonic Turbulence}
\shortauthors{Robertson \& Goldreich}

\newcommand\NST{N_{\mathrm{ST}}}
\newcommand\Mach{\mathcal{M}}
\newcommand\Mbar{\bar{\Mach}}
\newcommand\cs{c_{\mathrm{s}}}
\newcommand{\rhobar}{\bar{\rho}}
\newcommand{\rhothresh}{\rho_{\mathrm{thresh}}}
\newcommand{\tcol}{t_{\mathrm{col}}}
\newcommand{\tcross}{t_{\mathrm{cross}}}
\newcommand{\rhow}{\rho_w}
\newcommand{\rhomax}{\rho_\mathrm{max}}
\newcommand{\rhostar}{\rho_{\star}}
\newcommand{\vxi}{\vec{\xi}}
\newcommand{\vel}{\varv}

\begin{document}

\title{Dense Regions in Supersonic Isothermal Turbulence}



\author[0000-0002-4271-0364]{Brant Robertson}
\affiliation{Department of Astronomy and Astrophysics, University of California, Santa Cruz, 1156 High Street, Santa Cruz, CA 96054}

\author{Peter Goldreich}
\affiliation{California Institute of Technology, 1200 East California Boulevard, Pasadena, CA 91125}

\begin{abstract}

The properties of supersonic isothermal turbulence influence a variety
of astrophysical phenomena, including the structure and evolution of
star forming clouds. This work presents a simple model for the 
structure of dense regions in turbulence in which the density
distribution behind isothermal shocks
originates from rough
hydrostatic balance between the pressure gradient behind the
shock
and its deceleration from ram pressure applied
by the background fluid. Using simulations of supersonic
isothermal turbulence and idealized waves moving through a
background medium, we show that the structural properties
of dense, shocked regions
broadly agree with our analytical model.
Our work provides a new conceptual picture for describing the dense 
regions, which complements theoretical efforts to 
understand the bulk statistical properties of turbulence and attempts
to model the more complex features of star forming clouds like
magnetic fields, self-gravity, or radiative properties.

\end{abstract}

\keywords{hydrodynamics --- 
turbulence --- ISM: clouds --- stars: formation }

%
%
\section{Introduction}
\label{section:intro}

The physics of star formation and
molecular gas in galaxies depend on the
properties of supersonically 
turbulent clouds. Observed line widths
indicate the presence of supersonic,
random
bulk motions within interstellar clouds,
and a combination of collisional heating
and radiative cooling
keeps their gas roughly isothermal
despite any strong shocks
that develop.
Protostars can condense
from gravitationally bound regions within
cold molecular gas, and the supersonic
isothermal
turbulence within the bound clouds will persist
as long as collapse occurs faster than 
the largest turbulent eddies turn over
\citep{robertson2012a,murray2015a,murray2017a},
until magnetic fields become important \citep[e.g.,][]{hennebelle2008b,chen2014a},
until the gas becomes optically thick to its
own cooling radiation,
or the conversion
of gravitational potential or nuclear energy 
into kinetic energy disperses the cloud.
The properties
of dense turbulent clouds therefore 
set the initial conditions of the
star formation process on smaller scales,
and a deeper understanding of the
physics of dense regions in turbulence
will enable a more complete picture for
how interstellar clouds transform into
stars.

To this end, this paper develops a new theoretical
model for dense regions in supersonic
isothermal turbulence that explains their
internal structure and time evolution. Using
a combination of hydrodynamical simulation 
and new analysis methods, we identify the
population of dense regions, measure their
physical structure, and characterize their
features. Our
work connects the properties of individual
dense regions to the statistical properties of the
supersonically turbulent fluid, and
provides a new view for how gravitational collapse initiates.

The increasingly rich set of observations of 
molecular gas clouds acquired over the last
forty years provides a strong empirical 
motivation for modeling interstellar medium
(ISM) clouds as turbulent fluids. The
velocity-size relations of molecular clouds
\citep{larson1981a,myers1983a,solomon1987a,goodman1998a,bolatto2008a,heyer2004a,heyer2009a} 
finds an analog in the velocity structure  
function of turbulent motions
\citep{elmegreen2004a}. 
Other observed properties of molecular clouds,
such as their filamentary morphology in the radio
\citep{,schneider2011a,kirk2013a} and in {\it Herschel} infrared data
\citep{andre2010a,menshchikov2010a,miville-deschenes2010a,arzoumanian2011a,hennemann2012a,schneider2012a,konyves2015a},
or their approximately fractal
character \citep{stutzki1998a,roman-duval2010a},
suggest they contain supersonically turbulent
gas. Indeed, maps of molecular clouds resemble
the projected density fields of simulated
turbulent fluids \citep[e.g.,][]{federrath2010a,smith2014a}, 
with both possessing 
large spatial inhomogeneities \citep[e.g.,][]{falgarone1992a} 
and dense, filamentary features.

Simple analytical and dimensional
arguments provide deep reaching physical
descriptions of the properties of 
incompressible turbulence \citep{kolmogorov1941a},
and subsonic magnetohydrodynamical turbulence has a well-developed
analytical theory for how dissipation proceeds \citep[e.g.,][]{goldreich1995a,goldreich1997a}.
However, the shock-ridden
structure of supersonic turbulence
limits analytical models from providing a complete
picture. 
In contrast to the roughly local (in
$k$-space) interactions between vortices that
describes the energy cascade in 
incompressible turbulence \citep{kraichnan1959a}, 
the nonlocal
interactions between large-scale bulk motions
and dissipation occurring on small scales near
shocks
has mostly stymied rigorous analytical
modeling. 
For instance,
the velocity power spectrum of supersonic
turbulence is intermediate between Kolmogorov
and the Burger's spectrum for pure shock turbulence,
and may require a density-weighting to describe
approximately through analytical means \citep{kritsuk2007a,federrath2013b}.

This challenge has motivated the engineering
of sophisticated numerical simulations of
the properties of supersonic turbulence, through
which much of the current intuition about the
role of turbulence in molecular clouds has
been built. Simulations have verified that
random motions in supersonic turbulence dissipate
roughly on the Mach crossing time of the fluid,
without or without the presence of magnetic fields
\citep[e.g.,][]{stone1998a,maclow1998a,maclow1999a,ostriker2001a,cho2003a,beresnyak2011a}.
This finding suggests that turbulence in
real molecular clouds must be regularly driven or
the interior structure of the cloud will evolve
on a short time scale.
The velocity structure function of supersonic
turbulence shows a steep relation between velocity
differences and scale \citep{ballesteros-paredes2006a,kritsuk2007a},
similar to the size-line width relation
for molecular clouds, which may indicate that
clouds of different sizes have similar turbulent
properties. 

Connections drawn between models for supersonic turbulence
and the theory of star formation often involve
the statistical properties of the turbulent 
density field \citep[for reviews, see][]{maclow2004a,mckee2007a,krumholz2014a}. Supersonic isothermal turbulence
displays a volumetric density probability density
function (PDF) close to lognormal for solenoidally-driven 
turbulence
\citep{vazquez-semadeni1994a,padoan2002a,kritsuk2007a}.
The shape of the PDF has been ascribed to the statistics
of random, overlapping density modes \citep{vazquez-semadeni1994a,padoan2002a},
which emphasizes
the very statistical picture for understanding
astrophysical turbulence to date.
The width of the PDF depends on the turbulent
Mach number, such that the density contrasts
increase as the bulk motions become more
supersonic \citep[e.g.,][]{lemaster2008a}.
The morphology of density
inhomogeneities and the corresponding shape of
the density PDF also depend on whether the turbulent
forcing field is primarily solenoidal or
compressive \citep[e.g.,][]{federrath2008a,federrath2010a}, 
suggesting that the observed properties of 
molecular clouds may encode the nature of the
driving mechanism \citep[e.g.,][]{ginsburg2013a}.
In star-forming clouds the line-of-sight extinction
and inferred column density PDFs develop a power-law behavior
at high densities \citep[e.g.,][]{kainulainen2009a,arzoumanian2011a,schneider2012a},
a feature which has been reproduced by turbulence simulations that include self-gravity
\citep[e.g.,][]{kritsuk2011a,ballesteros-paredes2011a,lee2015a,burkhart2017a}.

These statistical properties of the turbulent
density field provide the elements for a relatively
simple picture of star formation in molecular clouds.
Supersonic turbulence within a cloud is generated
by a driving field, setting the velocity structure
and density inhomogeneities of the gas. The combination of
the velocity-size scaling relation with observed
correlations involving the cloud mass indicate that
gravitational potential and kinetic energies
of molecular clouds lie close to virial balance
\citep{larson1981a,solomon1987a,bertoldi1992a,krumholz2005a}.
Given the strength of gravity, virial balance sets
the largest scale on which the cloud is marginally
bound. The density PDF then indicates what fraction 
of the gas lies at densities above some Jeans-like
instability criterion, which sets the fraction
of gas that collapses via self-gravity \citep{krumholz2005a}.
The average efficiency of star formation in molecular clouds
is low \citep{krumholz2007a}, with typically a few percent of
the cloud mass converted to stars on a free-fall time scale. 
Observationally, star formation rates
scale with the abundance of molecular gas \citep{gao2004a,bigiel2008a,kennicutt2012a}
or the fraction of dense molecular gas \citep{lada2010a,lada2013a,evans2014a,lada2017a},
but there has been some disagreement about how that connection arises
physically \citep{lada2012a,krumholz2012a}. 

By choosing
the threshold for star formation appropriately and accounting for other
relevant properties of the turbulence (e.g., magnetic 
field strength), low star formation
efficiencies of a molecular cloud can be reproduced
\citep[e.g.,][]{krumholz2005a,padoan2011a,federrath2012a,kainulainen2014a,padoan2017a}.
Detailed simulations of 
star-forming clouds use similar criteria to determine
the regions that ultimately collapse into stars,
often by placing sink particles in potential
minima with converging velocity fields subject to
constraints on the proximity of infalling regions.
These models for star formation in molecular clouds
enjoy considerable success in matching the observations
of star-forming regions and the resulting population
of dense cores and stars
\citep{klessen1998a,klessen2000a,klessen2001a,bate2003a,bonnell2003a,bonnell2006a,glover2007a,glover2007b,krumholz2007a,offner2008a,girichidis2011a,federrath2012a,federrath2013a,federrath2015a,liptai2017a,haugbolle2017a}, although the relative importance
of driving mechanisms, feedback, initial cloud structure, magnetic fields, or
other physics remains unclear.

Despite the successes of these models, some 
important puzzles still remain in relating
isothermal, supersonically-turbulent fluid to 
a real star-forming cloud. If the cloud persists
over long time scales \citep[e.g.][]{blitz1980a},
the large-scale forcing of the cloud turbulence must 
operate repeatedly on time scales less than the
Mach crossing time. 
For simulations
where the turbulence has reached steady statistical
state, the forcing has typically 
been applied many times over.

If turbulent
motions marginally support the cloud
against self-gravity on large scales,
as the apparent virial balance may imply,
then
the bulk of the cloud
might survive as long as a source of
regular driving remains available.
Under such conditions,
the density structure of the turbulence
within the cloud will give rise to regions
that will nonetheless collapse on time scales substantially
shorter than the Mach crossing time of the whole
cloud. These dense interior regions will form
stars once they collapse, and several outcomes
are possible. If the gravitational potential
or nuclear energy can be converted into kinetic
energy through the star formation process (i.e.,
feedback), then the 
star formation itself could in principle drive
the cloud turbulence \citep{maclow2004a,federrath2015a}. 
However, to prevent the
collapse of the whole cloud the forcing has to
be applied on large scales and coupled to gas
throughout \citep[e.g.,][]{vazquez-semadeni2003a,brunt2009a}. If the feedback can
be efficiently coupled to the gas, then substantial mass
from the marginally bound cloud could be 
freed.
If the feedback cannot sustain the turbulence
but does not dissipate the cloud, then a persistent
cloud would again require continuous external driving
and perhaps a steady inflow of gas to balance its
star formation rate.
Otherwise, the star-formation efficiency becomes
time-variable and increases as the molecular clouds
disrupt \citep{murray2011a}.

The difficulties in arranging a long-lived
turbulent cloud with successive generations of
star formation have motivated models beyond
the simple turbulent box picture. Converging
flows can drive turbulence and lead to realistic 
molecular clouds \citep[e.g.,][]{ballesteros-paredes1999a,heitsch2005a,vazquez-semadeni2006a,heitsch2008a,heitsch2009a,heitsch2011a,chen2014a,kortgen2015a,kortgen2017a,inoue2017a}. Clouds can
be continually formed during the time scale of
the converging flow, but their turbulence will
decay on a Mach crossing time once the
large-scale convergence ends. Unless the
convergence is somehow permanent or another large-scale
driving mechanism is created (see above), the clouds
will eventually undergo a rapid end where dense
bound regions will convert to stars and the cloud
will dissipate on large scales, perhaps owing to
feedback.

In a picture where star-forming molecular clouds experience
short lifetimes comparable to or less than their Mach crossing
times, the original formation of the cloud would
need to generate its interior turbulent structure.
Once regions within the cloud become overdense enough
to become gravitationally bound, the evolution of the
cloud proceeds quickly. Bound regions form stars, and
the short-lived massive stars provide feedback energy
to the surrounding gas that may affect the overall
cloud star formation efficiency but does not supply
effective large-scale driving to sustain the cloud
turbulence over the long term. The cloud may be
dispersed owing to the star formation feedback
as the turbulence decays, the 
kinetic energy in bulk motions dissipates, and the
density inhomogeneities reduce. Star formation on 
large scales within a galaxy would be connected to
the rate at which molecular clouds form, through
converging flows \citep[e.g.,][]{hartmann2001a,dobbs2008a}, large scale gravitational instability,
or other means, and the processes that set the
star formation efficiency of the clouds as regions
within them collapse \citep[e.g.,][]{braun2015a,semenov2016a}.

A model for long-lived molecular clouds could assert
that the observed cloud velocity-size relations
result from all clouds maintaining a
marginal virial balance, sustained by a persistent
driving mechanism.
Short-lived molecular cloud models still must
reproduce the observed cloud scaling relations, but
cannot rely on replenishment of the turbulent
motions from large-scale driving. 
The nature of the
gravitational collapse itself has to maintain the
observed scaling relations by driving turbulence
\citep[e.g.,][]{scalo1982a,ballesteros-paredes2011a,ibanez-mejia2016a}. 
In \citet{robertson2012a},
we identified how collapsing regions
undergo ``adiabatic heating'' of the turbulence
if the collapse occurs quickly
compared to the initial Mach crossing time. We showed
how eventually
the collapse rate and the large-scale eddy turnover
rate in the cloud will synchronize, leading to a
connection between the turbulence within the cloud
and its gravitational collapse, and suggested
the size-dispersion relation for clouds reflected
this connection. \citet{murray2015a} showed that
adiabatic heating during gravitational
collapse can explain changes in the size-line width
relation in massive star-forming regions \citep{fuller1992a,caselli1995a,plume1997a}. They showed that, in the presence of adiabatic
heating, within the sphere of
influence of a collapsing region the turbulent velocity
increases with decreasing radius. This feature contrasts
with earlier models of collapse where the
character of turbulent velocities during infall did not
change \citep{mckee2003a}.
In simulations of
turbulent self-gravitating gas \citet{murray2017a} showed that the
turbulent velocities increase with decreasing radius during
the gravitational collapse as $\vel\propto r^{-0.5}$, as we
speculated in \citet{robertson2012a}.
Other recent simulations of star-formation in turbulent
gas show consistency with the \citet{murray2015a}
model \citep[e.g.,][]{mocz2017a,ibanez-mejia2017a,li2017a} for the structure
of self-gravitating regions shaped by adiabatic heating.

If the gravitational collapse of turbulent clouds
proceeds in a manner that can reproduce the size-line width
relations, then the picture forwarded by \citet{murray2015a}
of molecular
clouds as a collapsing turbulent flow appears viable.
A remaining issue for
this model is how the star formation efficiency
connects with the internal structure of the cloud.
Therefore, understanding the properties of dense 
regions in supersonic turbulence including
their density profiles, turbulent lifetimes and
structural evolution, spatial clustering,
connection with the gravitational potential,
and relation to the statistical properties of
the turbulent medium is
of interest. 

Below, we present the
results of supersonic isothermal turbulence
simulations where we have characterized in
detail the properties of dense regions. 
Section \ref{section:simulation}
describes how our hydrodynamical turbulence
simulations were performed. 
Section \ref{section:dense_regions} presents
our method for identifying dense regions
and measurements of their individual properties.
We develop an analytical model for their
internal density structure based on 
exponential isothermal atmospheres in
Section \ref{section:exponential_atmospheres}.
The time-dependent properties of the dense
regions are studied in Section \ref{section:time_dependence},
including a measurement of the typical lifetimes
of the densest regions in Section \ref{section:shock_lifetimes}. 
The spreading of shocked regions in response to 
deceleration from on-coming ram pressure is examined in
Section \ref{section:spreading}.
We compute the
collective properties of
the population of dense regions in 
Section \ref{section:shock_populations},
including spatial clustering (Section \ref{section:clustering})
and their contributions to the density
PDF (Section \ref{section:dense_pdf}).
We then consider how the gravitational
potential of the turbulent cloud might
affect the dense regions in Section \ref{section:gravity}.
A discussion of our results is presented
in Section \ref{section:discussion}, along
with our conclusions in \ref{section:conclusions}.
A host of analysis methods
were engineered for studying the properties of
dense regions in turbulence, and these methods
are described in more detail in a set of Appendices.
Throughout the paper, we will refer to the dense fluid structures bounded by shock discontinuities as ``shocked regions''. The terms ``pre-shock'' and 
``post-shock'' will indicate areas ahead and behind of a shock, respectively.

%
%
\section{Turbulence Simulations}
\label{section:simulation}

\begin{figure*}[ht!]
\begin{center}
\includegraphics[width=7.05in]{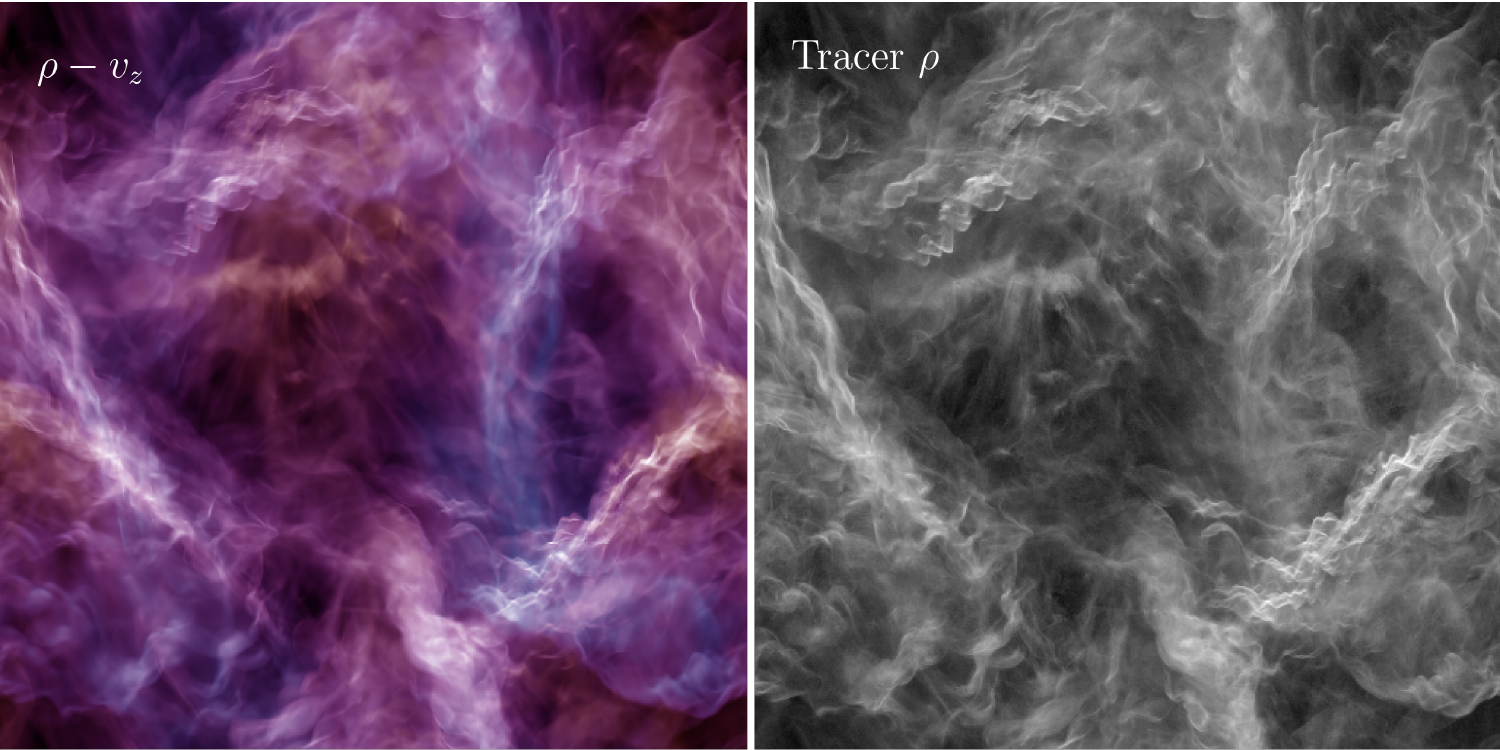}
\caption{Hydrodynamical simulation of supersonic isothermal turbulence. Shown are a
logarithmic projection of the average turbulent fluid density integrated through the $N=512^3$ grid (left panel; $0.25<\langle\rho\rangle<4.5$) colorized by the
vertical velocity ($0<\vel_z<10$ in red; $-10<\vel_z<0$ in blue), and a logarithmic projection of the
number of tracer particles evolved with the fluid (right panel).
\label{fig:box}}
\end{center}
\end{figure*}

To study dense regions in turbulent clouds, 
we perform
simulations of supersonic isothermal turbulence using a modified
version of the hydrodynamics
code {\it Athena} \citep{stone2008a}.  The simulations follow
the calculations presented in \citet{robertson2012a},
with a few modifications.  The calculations simulate
an isothermal fluid (with sound speed $\cs=1$) in a unit box (side length $L=1$)
with mean density $\rhobar=1$, evolved
on either $N=512^3$ or $N=1024^3$ grids
using linear reconstruction and a
constrained transport upwind integrator \citep[see][]{colella1990a,gardiner2008a}. 
Following \citet{kritsuk2007a}, an
acceleration field generated with a flat spectrum with power only in the
first two $k$-modes drives the fluid. The driving field is constrained
to be solenoidal
by performing a Helmholtz decomposition in Fourier space on a generic field produced
from an appropriate transfer function applied to white noise, using the method described 
by \citet{bertschinger2001a}. 
The forcing field is applied ten times per crossing
time $\tcross = L/(2\Mbar \cs)$ with an amplitude chosen to maintain a root-mean-squared (RMS)
Mach number of $\Mbar\approx5$.  

The $N=512^3$ simulation is run for fifty crossing times, and the
conserved quantities from the simulation grid are recorded ten times per crossing time.
After twenty-five crossing times, the simulation is output at a rate of
5000 snapshots per crossing time over a brief duration of one tenth of a crossing time.  
Afterward, the simulation data are 
again saved ten times
per crossing time.  During the last twenty crossing times, the forcing is turned off
and the turbulence is allowed to decay. The resulting $\sim1300$ snapshots provide a wealth of
information on the time-dependent properties of supersonic turbulence.

For the $N=1024^3$ simulation, we drive the turbulence continuously to
maintain an RMS Mach number of $\Mbar\approx5$ and perform our analysis
on a single snapshot output after four crossing times. While this higher
resolution simulation is driven by different realizations of the forcing
field than is the $N=512^3$
simulation, we have checked that the statistical
properties of both simulations are consistent. We use the results of the
$N=1024^3$ simulation to verify that our conclusions are insensitive to
resolution, as discussed in Section 4 below.

The left panel of Figure \ref{fig:box} shows a visualization of the entire 
$N=512^3$ simulation volume
after twenty turbulent crossing times. The 
image intensity is scaled with a logarithmic
projection of the density through the simulation, while the coloration reflects whether the
average projected fluid velocity in the vertical direction is positive (red) or negative (blue).
The classic features of supersonic turbulence are apparent, with
large density inhomogeneities in the fluid spanning $\sim6$ orders of magnitude
in the range $10^{-3} \lesssim \rho/\rhobar \lesssim 10^3$. The main focuses of this paper are
the structural properties and evolution of the dense regions, which appear bright white in
Figure \ref{fig:box}.

To assist in our analysis of dense regions in turbulence, we have implemented a new tracer
particle scheme into {\it Athena}. The details of this numerical scheme are presented in
Appendix \ref{section:tracers}. The tracer particles are initially distributed with the
grid, but move in response to the fluid velocity interpolated from the grid. Throughout the
paper, we use the tracer particles to define dense regions, track their evolution with
time, measure the statistical properties of the population of dense regions, 
and connect the dense regions to the gravitational potential that the turbulent
gas would generate given its density structure.
Figure \ref{fig:box} shows the number of tracer particles projected through the
$N=512^3$ simulation volume, scaled logarithmically (right panel). Very similar density inhomogeneities
are apparent in both the fluid simulated on the grid and
tracer particles. The tracer particles do not represent Lagrangian mass elements \citep[e.g.,][]{genel2013a},
but do provide convenient locations for measuring approximate fluid properties interpolated
from the grid. The particle interpolation methods are discussed in detail in Appendix \ref{section:reinterpolation}.

%
%
\section{Dense Regions in Turbulence}
\label{section:dense_regions}

\begin{figure*}[ht]
\includegraphics[width=7.1in]{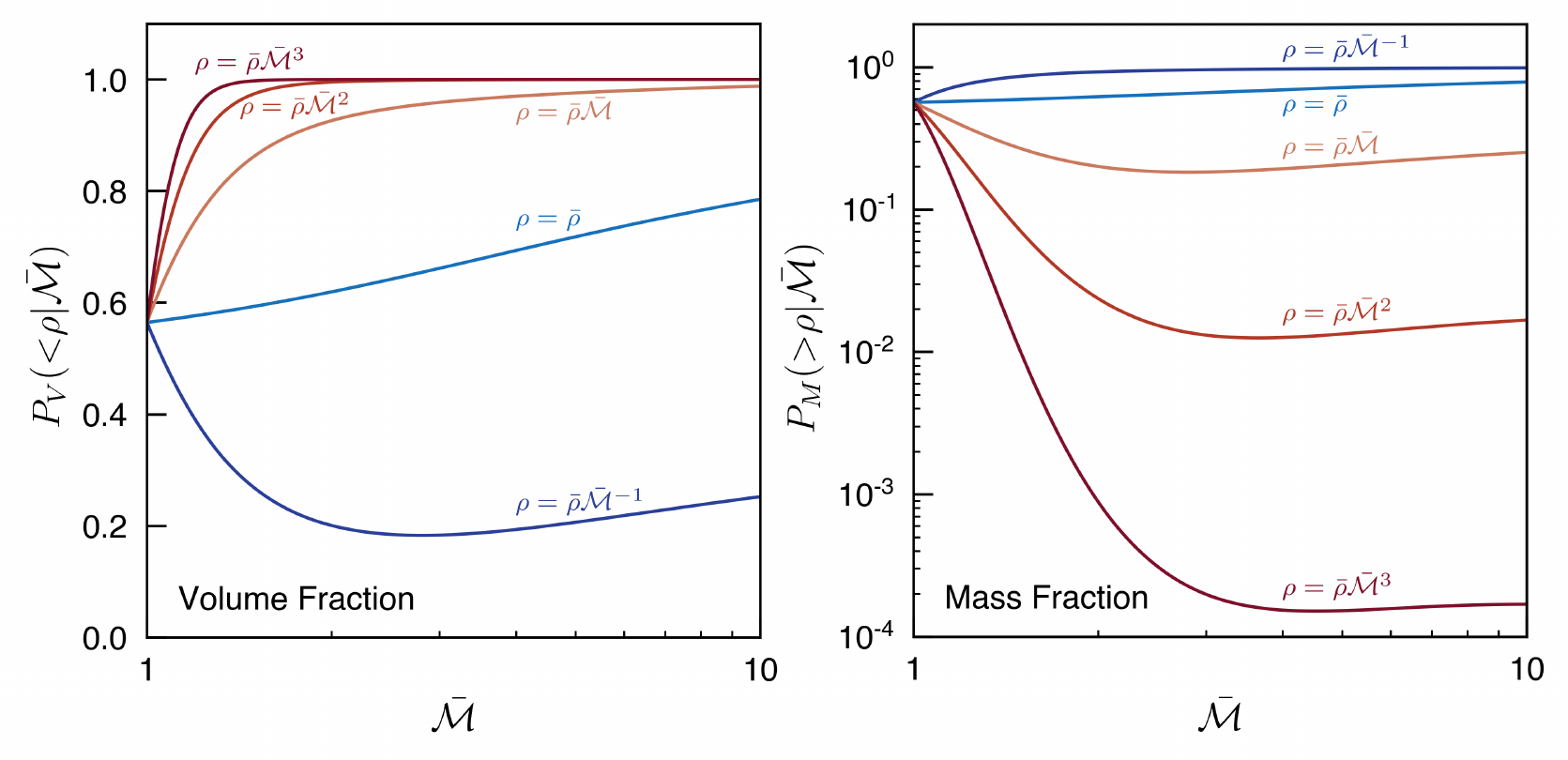}
\caption{Volume (left) and mass (right) fractions
of supersonically turbulent isothermal fluid below and above a given density, respectively. Shown
are the volume and mass fractions, defined in Equations \ref{eqn:volume_fraction} and
\ref{eqn:mass_fraction} for density thresholds $\rho/\rhobar = [1/\Mbar,1,\Mbar,\Mbar^2,\Mbar^3]$. Most of the volume in supersonic isothermal turbulence lies at densities
$1/\Mbar\lesssim\rho/\rhobar\lesssim 1$, while
most of the mass resides in regions with
densities $1\lesssim\rho/\rhobar\lesssim\Mbar$.
\label{fig:mass_fractions}}
\end{figure*}

The simulations described in Section \ref{section:simulation} reproduce the well-known
phenomenologies of supersonic isothermal turbulence
studied extensively in the literature \citep[e.g.,][]{kritsuk2007a,federrath2010a}. The velocity power
spectrum is steeper than Kolmogorov, with the
high-frequency power-law behaving as $P(k)\propto k^{-\alpha}$ with $\alpha\approx 1.7-1.9$ depending on
time variations. In agreement with previous
work, the volumetric PDF of density $\rho$
for our solenoidally-driven simulation
is close to a lognormal of
the form
\begin{equation}
\label{eqn:pdf}
p(x|\Mbar)dx = \frac{1}{\sqrt{2\pi\sigma^2(\Mbar)}}\exp\left[-\frac{(x-\mu)^2}{2\sigma^2(\Mbar)}\right]dx,
\end{equation}
\noindent
with $x\equiv \log \rho/\rhobar$,
a dispersion $\sigma$,
and the constraint
$\mu = -\sigma^2(\Mbar) / 2$. Previous authors have
found that the dispersion scales with the root-mean-squared (RMS)
turbulent Mach number
$\Mbar$ as
\begin{equation}
\sigma^2(\Mbar) = \log\left(1 + b^2 \Mbar^2\right),
\end{equation}
\noindent
where the constant $b\sim0.2-0.5$
\citep[e.g.,][]{padoan1997a,passot1998a,li2003a,kritsuk2007a,lemaster2008a,federrath2010a,price2011a,konstandin2012a,molina2012a}.
In what follows, we will distinguish between
the RMS Mach number $\Mbar$ that describes the
typical bulk random velocity of fluid in the
turbulence, and the Mach number $\Mach$ of 
individual shocks.

Some implications of the density PDF on the 
formation and 
evolution of dense regions in turbulence can be foreseen from integrals of Equation \ref{eqn:pdf}, as shown in Figure \ref{fig:mass_fractions}. Displayed are the volume integrals
\begin{equation}
\label{eqn:volume_fraction}
P_V(<\rho | \Mbar) = \int_{-\infty}^{\ln \rho/\rhobar} p(x|\Mbar) dx
\end{equation}
\noindent
and the mass integrals
\begin{equation}
\label{eqn:mass_fraction}
P_M(>\rho | \Mbar) = \int_{\ln \rho/\rhobar}^{\infty} p(x|\Mbar) e^x dx
\end{equation}
\noindent
indicating the fraction of the
volume below and the mass above densities
$\rho/\rhobar$, as a function of the RMS Mach
number $\Mbar$. The volume-filling densities lie at
$1/\Mbar\lesssim \rho/\rhobar \lesssim 1$, while most
of the mass has densities $1\lesssim\rho/\rhobar\lesssim \Mbar$. These fractions are only weakly dependent on $\Mbar$, and a rough rule of thumb is
that for solenoidally-driven turbulence 
the mass fractions $P_M(\rho>\Mbar^\alpha \rhobar | \Mbar) \lesssim \Mbar^{-\alpha}$. Turbulence driven
with compressional modes deviates from the lognormal
PDF, and can have somewhat higher
velocity- and mass-fractions in dense regions \citep[e.g.,][]{federrath2008a}.

The rough factor of $\Mbar^2$ between the volume-filling density and the mass-occupying
density is not accidental, and arises from the
compression factor $\Mach^2$ for isothermal
shocks.
By design, most of the volume and
mass of fluid in the simulation move with
relative velocities $\vel\sim\Mbar\cs$. This connection
gives rise to the concept of a {\it first-generation}
shocked region
in turbulence, generated by encounters
between regions with the volume-filling density
$\rho\sim \rhobar/\Mbar$ at the typical relative
velocity $\vel\sim\Mbar\cs$.

Regions with densities $\rho\gg\Mbar\rhobar$
occupy very small volumes ($\lesssim$1 percent)
and comprise a
small fraction of the total mass of the
fluid (a $\sim$few percent) in supersonic
isothermal turbulence. Often, the vast majority of
computational effort in these simulations 
is therefore spent elsewhere, on either the
regions with volume-filling or mass-occupying
densities. The statistical measures typically
applied to turbulence simulations, such as the
velocity power spectrum, are volume-weighted and
therefore largely ignore the densest regions
in turbulence. 

Since dense regions occupy such a small volume,
chance encounters between dense regions are
relatively rare. If it survives long enough,
a given region with $\rho_1\gg\rhobar$ could
travel a significant fraction of the simulation
volume without colliding with another region with
$\rho_1>\rho_2\gg\rhobar$ if the densities are
comparable (e.g., $\rho_1/\rho_2<\Mbar$).
This fact bears on whether very dense regions
are produced as {\it higher-generation} 
shocked regions,
meaning they are produced 
through generations of collisions between
shocks
traveling at velocities $\vel\sim\Mbar\cs$,
or whether they are {\it high-velocity} 
shocked regions
where a large relative velocity between the
pre- and post-shock regions give rise to a
very large density contrast. We discuss this
issue in more detail below.

\subsection{Measuring the Properties of Dense Regions}
\label{subsection:dense_regions}

Dense regions occupy small fractions
of the volume and mass of a turbulent fluid.
The three-dimensional structure of turbulence
is famously complex, and identifying and
characterizing the properties of the
densest regions requires additional analysis
effort beyond performing the simulation itself.
Figure \ref{fig:box} illustrates the complexity
of identifying distinct dense regions in turbulence,
as dense structures, which appear as filaments
in projection, seemingly overlap and do not
clearly exist as individual ``objects'' \citep[e.g.,][]{smith2016a}.
This complexity
arises in part because dense regions are 
bounded by shocks,
and
are generated in the interaction of waves in the fluid
that have a wide extent in frequency space. The
projection of the density field also implies connections
between regions along the line of sight,
but in many cases these regions can be separated by
surrounding regions of substantially lower densities.
Nonetheless, the density field appears complex and 
some methodology for identifying individual
shocked regions
needs
engineering.

The problem of identifying individual dense structures in
supersonic turbulence is not unlike the task of 
cataloging dark matter halos in cosmological N-body
simulations \citep[see, e.g.,][]{knebe2011a}, with
some notable differences. The complexity of the
density field in turbulence leads to the ``cloud-in-cloud''
problems encountered when identifying substructure 
during halo finding, except with actual clouds.
In the absence of self-gravity, turbulence does not
have a virial condition to define the extent of
regions of interest. Further, in the absence of
self-gravity, regions in turbulence are not
Lagrangian features. Indeed, the densest regions
in turbulence are shocks and material may
pass from the pre-shock to the post-shock regions ahead
and behind of the shock quickly. The intermittency
of turbulence suggests that the properties of
dense regions may themselves change on relatively
short time scales
\citep[e.g.,][]{klessen2000a,vazquez-semadeni2005a,glover2007a}, and further complicates
the analysis of dense regions in turbulence.

To study dense regions, we therefore
require methodologies
for identifying, measuring, and following them
over time. 
We have engineered some new techniques
for accomplishing these tasks, and present those
methods in Appendices \ref{section:group_finding},
\ref{section:shock_orientation},
and \ref{section:group_tracking}. The key issues
in developing these algorithms include separating
distinct regions in the density field, defining
a natural frame-of-reference for dense regions
that often involve velocity shifts and rotations
from the simulation frame and coordinates, and
the time-tracking of non-Lagrangian regions whose
particle content can evolve over short time scales.
These issues do not have unique solutions, but
our methods resolve them satisfactorily for the
purposes of this work. We refer the interested
reader to the Appendices for more detail. Depending
on the time step, we typically identify several
thousand independent regions with densities
$\rho\gtrsim \Mbar^2\rhobar \sim 25\rhobar$. For
simulations with $N=512^3$ tracer particles, the
dense regions contain $10-10,000$ particles 
at $\rho\ge25\rhobar$ depending on each region's
peak density $\rho_0\approx (25-300)\rhobar$. 
We now turn to applying the techniques
we have engineered for measuring the properties
and time-evolution of these dense regions in
supersonic turbulence.

%
%
\section{Shocked Region Profiles}
\label{section:shock_profiles}

A prominent feature of isothermal shocks is the $\Mach^2$-contrast in pre- and post-shock densities, 
and for normal shocks of infinite extent this relation inferred from the Rankine-Hugoniot conditions provides
a complete description of the density structure of the flow near the shock
\citep[e.g.,][]{shu1991a}.
The post-shock structure behind real isothermal 
shocks
are
not solely specified by the jump condition, and as is apparent from Figure \ref{fig:box} the
individual shocked regions
are quite thin with large negative density gradients behind the shock. Using our 
methods for identifying and measuring the properties of dense regions, we can determine the structure
of individual shocked regions
and develop a physical model for their density profiles.

\begin{figure*}[ht]
\begin{center}
\includegraphics[width=7.1in]{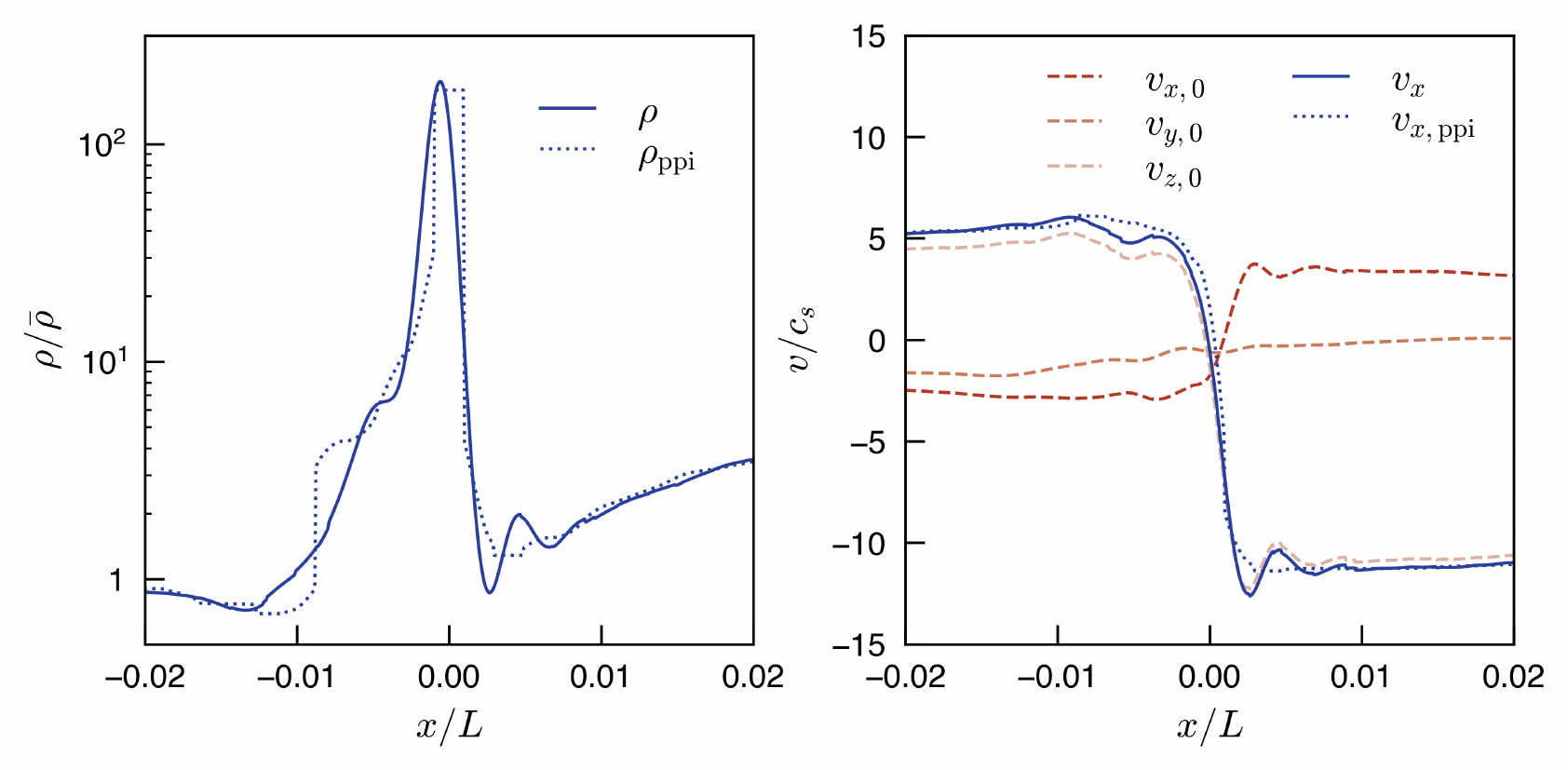}
\caption{\label{fig:example_shock} Density (left) and
velocity (right) profiles for an example shocked region
identified in a 
supersonic isothermal turbulence simulation. The shocked region
 was
identified from a peak among the tracer particles in the simulation,
and the tracers were used to identify the orientation and direction of
travel in the simulation volume. The $x$-direction indicates the direction
of travel that is primarily orthogonal to the shock front, while the
$0$-subscripts indicate coordinates aligned with the simulation volume
reference frame.
In both panels, Gaussian process (solid lines) and piecewise parabolic (dotted
lines) interpolations through the simulation volume are shown. The
large density
contrast of the shock
(left panel) relative to the mean density $\rhobar$ results
from its high Mach number ($\Mach = \vel_x/c_s \approx 15$), as the pre-shock
density is only $\rho\approx\rhobar$. The post-shock density profile appears roughly
exponential. This example shocked region
is nearly aligned
with the $z_0$-direction of the simulation volume, and has a primary
direction velocity $\vel_x\approx \vel_{z,0}$ (solid line, right panel).
The velocities of the shocked region
in the simulation box reference frame are shown as dashed lines in the right panel. The abscissae in both panels are scaled relative to the simulation box size $L$.
}
\end{center}
\end{figure*}

Figure \ref{fig:example_shock} shows the density and velocity field near an example shock with
peak density $\rho_0 \approx 230 \rhobar$. The one-dimensional profiles are centered about the local
peak in the density field and oriented using information from
the moment of inertia tensor and the velocity field in the region.
Piecewise parabolic (PPI) and Gaussian
process (GPI) interpolations of the fluid properties are shown. The
``0'' subscript denotes the coordinate system of the simulation, and
the $x$-direction denotes the primary direction of travel of the 
shock.
This example
shocked region
is oriented near the $z_0$-axis of the simulation volume, such that the bulk
velocity of the shocked region
is nearly aligned with the $z_0$-direction.

In this example, the pre-shock density is close to the mean density and the
large density contrast relative to the mean
is primarily driven by the $\Mach^2$ change in the $x$-velocity
across the shock. This example is therefore a ``high-velocity'' 
shocked region.
 The
post-shock density profile eventually declines to near the mean density.
As is highlighted by the log-linear 
scale shown in Figure \ref{fig:example_shock}, the post-shock
density profile appears roughly exponential.

\begin{figure*}[ht]
\epsscale{1.1}
\plottwo{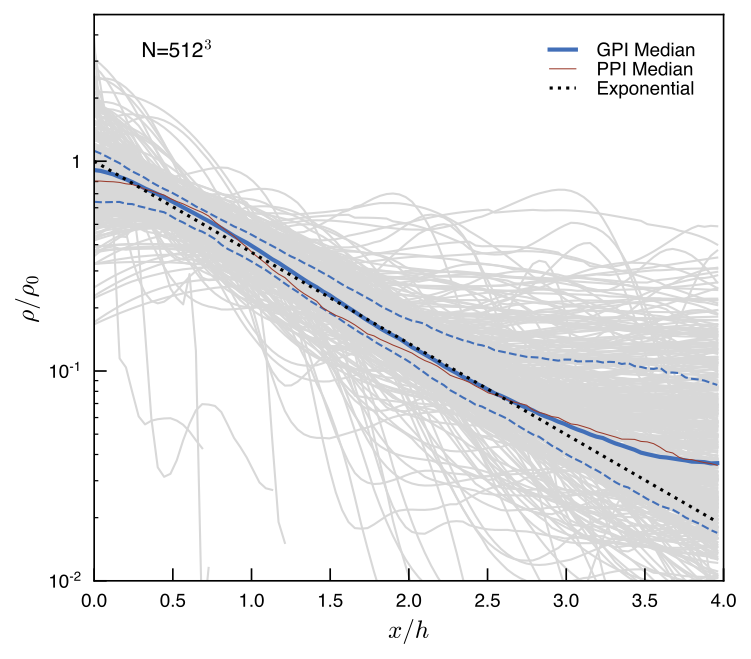}{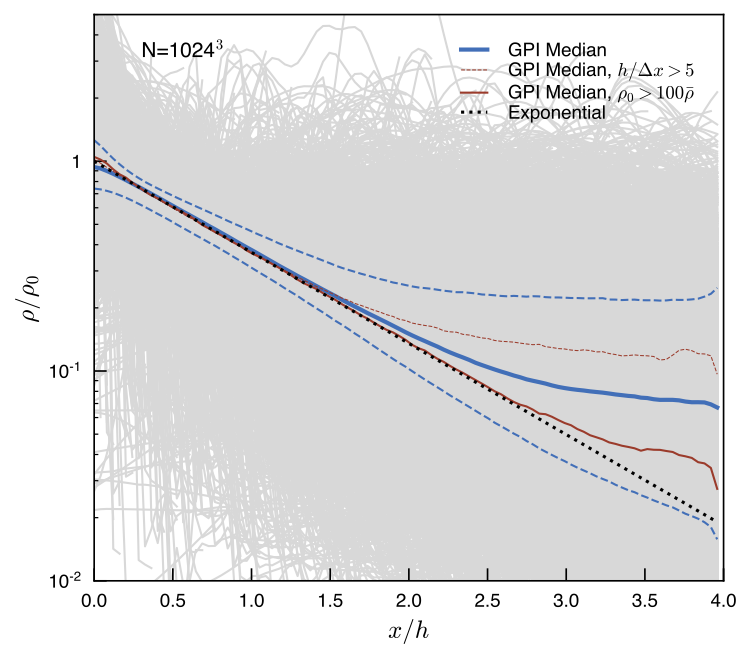}
\caption{\label{fig:density_profiles}
Post-shock density profiles of shocked regions
in simulations of supersonic isothermal
turbulence.
The left panel shows the individual density profiles of hundreds of 
shocked regions with peak densities $\rho_0>25\rhobar$
identified in the $N=512^3$
simulation using the tracer particles to separate distinct density enhancements.  A path through
each shocked region
is determined by information from its moment of inertia tensor and velocity field.
The post-shock region of the density profile measured along this
path is fit with an exponential $\rho\propto\exp(-h/x)$, where positive $x$ corresponds to 
a post-shock distance from the peak density.
Each profile is then rescaled by the scale length $h$ of the exponential, normalized by the maximum of
the exponential fit, and then plotted (gray lines).  At each position $x/h$ along the axis the
median (solid blue line) and inner 68\% spread (dashed blue lines) of
the Gaussian process interpolation (GPI) profiles can be measured, and 
compared with a exponential function (dashed black line). The corresponding median profile determined
from piecewise parabolic interpolation (PPI) profiles is shown for comparison (thin red line), using the exponential scale length and amplitude 
determined by fitting to the GPI profiles to rescale each PPI profile.
The comparison demonstrates that the median post-shock
density profiles is close to exponential out to a distance of at least $x/h\approx3$, even
as individual shocked regions
can show substantial deviations and the density profiles of some 
shocked regions
are poorly resolved.
The right panel shows the individual shock profiles for $\sim15,000$ shocked
regions identified in the $N=1024^3$ simulation (gray lines) with peak densities
$\rho_0>10\rhobar$. These regions also show exponential profiles (median
and inner 68\% spread shown as blue lines),
but encounter the surrounding background density at smaller $x/h$ than the
higher peak density regions shown in the left panel. In the
$N=1024^3$ simulation, a
larger number of the post-shock regions
are well-resolved ($h/\Delta x>5$; red dotted line).
These well-resolved regions typically 
have peak densities of $\rho_0\sim10\rhobar$,
and encounter the background after $x/h\approx1.5-2$. In contrast, the
densest peaks with $\rho_0>100\rhobar$ (thick red line)
show exponential profiles to $x/h\approx3$
before encountering the background density.
A physical model for the origin of the exponential profiles shown in 
both panels is discussed in 
Section \ref{section:exponential_atmospheres}.
}
\end{figure*}

\subsection{Average Density Profiles}
\label{section:density_profiles}

Given our method for identifying dense regions from the tracer particle distribution,
repeating the measurement illustrated in Figure \ref{fig:example_shock} 
for each shocked region
identified in the simulation is straightforward. Information from
the moment of inertia tensor defined by the tracer particles associated with each
shocked region
and their nearby velocity fields can be used to determine the 
shocked regions'
spatial
orientations. The trajectory of each 
shocked region
defines a skewer through the simulation
volume oriented roughly perpendicular to the associated shock
face, and the properties
of the simulated fluid can be interpolated along this skewer using the same
interpolation scheme used for assigning properties to the tracer particles.
Motivated by the roughly exponential post-shock density profile apparent in the
example shocked region
shown in Figure \ref{fig:example_shock}, we can fit exponentials to
the post-shock density profiles of each shocked region
and rescale them by their best-fit
amplitudes and scale lengths to place them on the same graph.

The left panel of Figure \ref{fig:density_profiles} shows
the ensemble of density profiles behind the five
hundred densest shocked regions
identified in a snapshot of
the $N=512^3$ simulation. Each shocked region
profile is rescaled by its fitted scale length $h$ and normalized by
the peak of the exponential fit, then plotted as a gray line. 
At each location $x/h$, the distribution of density profiles
can be measured.  The median (solid blue line) and inner 68\% variation (dashed blue line) of the
GPI density profile distribution is plotted in Figure \ref{fig:density_profiles}, along with
an exponential function (dotted black line). The median of the PPI density profiles is shown
for comparison as a thin red line, and is rescaled by
the GPI profile exponential fit amplitude and scale length parameters.
We find that the
median post-shock profile of these dense regions is very close to exponential out to at least $x/h\approx 3$.
The inferred scale lengths vary widely, from poorly- ($h\sim\Delta x$) to well-resolved ($h\gtrsim6\Delta x$).
Individual shocked regions
do show substantial variations from the exponential profile. Some
shocked regions
are clearly unresolved, and resemble an early solution to the isothermal two-shock Riemann problem with little difference between the pre- and post-shock profile shape (e.g., sharp discontinuities on both sides). Other 
shocked regions
can show exponential post-shock behavior out to roughly five scale lengths. More typically, shocked regions
in the simulations follow roughly exponential behavior in their post-shock density profiles for a few scale lengths and then show more complicated
density (and velocity) structure well behind the 
shock as the density profile approaches the average background density.

To further illustrate the exponential density profiles in shocked regions,
we can use the $N=1024^3$ simulation to study post-shock structures. The
right panel of Figure \ref{fig:density_profiles} shows $15,000$ density profiles
of shocked regions with peak densities $\rho_0>10\rhobar$ identified 
in the higher resolution simulation (gray lines). The previous fitting
procedure is
repeated, with the resulting median and inner 68\% spread in the GPI
density profiles shown as blue solid and dashed lines, respectively. These
lower density shocked regions show exponential behavior out to $x/h\approx1.5$,
at which point the profiles begin to encounter the background density
of the surrounding fluid. Restricting to shocked regions with density
profiles resolved with $h/\Delta x>5$ (thin red line) selects out shocked
regions with peak densities of $\rho_0\sim 10\rhobar$, which typically
encounter the background density by $x/h\approx1.5-2$. 
This measurement demonstrates that
restricting the analysis to well-resolved shocked regions
does not substantially change the median exponential behavior.
Restricting to the densest shocked regions with $\rho_0>100\rhobar$
(thick red line) extends the exponential behavior to $x/h\approx 3$,
similar to the densest regions examined in the $N=512^3$ simulation
(Figure \ref{fig:density_profiles}, left panel). The $N=512^3$ and
$N=1024^3$ simulations therefore find good agreement for the
typical density profiles of shocked regions.

%
%
\subsection{Exponential Atmosphere Model for Isothermal Shocked Regions}
\label{section:exponential_atmospheres}

The post-shock density profiles of shocked regions
measured in Section \ref{section:shock_profiles}
typically show a roughly exponential decline. This rapid fall-off of the density distribution
can be modeled using a physical picture for the formation and evolution of the isothermal
shocked regions
forming in the turbulence. In what follows, we present a physical model to explain
the general features of dense shocked regions
in isothermal supersonic turbulence based on exponential
atmospheres.

In turbulence simulations like those studied here, low frequency velocity perturbations are
introduced to drive large scale motions of the fluid and resupply energy into the turbulent
cascade. These perturbations can lead to substantial velocity variations in the fluid that
are compressive on small scales. Large compressive velocities between regions of typical densities
can result in high Mach-number shocks.

Initially, these shocked regions
can be extremely thin and display
sharp density contrasts (unresolved discontinuities) on either side of the density peak. Such regions
resemble the initial stages of a two-shock
isothermal Riemann problem, where the shock
conditions
would enforce a $\Mach^2$ density jump relative to the pre- and post-shock regions (with roughly constant densities and velocities) that comprise the local flow.
If the local flow were one-dimensional, this shock
structure would persist and the width of the dense
region
would simply increase as the forward and reverse shocks
moved into the pre- and post-shock regions.
However, given the complexity of the turbulent flow, the pre- and post-shock regions will have density and velocity structure such that the initial pressure balances generating the discontinuities
on either side of the dense region
will be upset. The density distribution in the post-shock region will re-adjust to
accommodate the pressure imbalance, with adjustments occurring over a sound-crossing time
across the narrow region. Provided that the original Mach number of the shock
is large, material
from the pre-shock region with density $\rho_w$ 
will still be encountered at a high relative velocity $\vel_w \approx \Mach \cs$.
The post-shock density profile of this region will necessarily adjust to provide a pressure
gradient $\nabla p$ that can reach 
hydrostatic balance with the decelerating force $\rho g$ owing to ram pressure $\rho_w \vel_w^2$
exerted by this
on-coming material.  We can 
describe this scenario mathematically by balancing the pressure 
gradient behind the shock
(of density $\rho$) with the ram 
pressure applied to the shocked region,
and write
\begin{equation}
\nabla p = - \rho g = - \rho \frac{\rho_w \vel_w^{2}}{\Sigma}
\end{equation}
\noindent
where $\Sigma = \int \rho dx$ is the mass per unit area of the
shocked region
measured along the
$x-$direction of travel. Writing $p = \rho \cs^{2}$ we have that
\begin{equation}
\label{eqn:hydro_equil}
\frac{d\rho}{dx} = -\rho \frac{\rho_w \vel_w^{2}}{\cs^{2} \Sigma},
\end{equation}
which gives the exponential solution $\rho(x) = \rho_{0} \exp \left( - x/h\right)$ with 
\begin{equation}
\label{eqn:scale_length}
h \equiv \frac{\cs^2}{g} =  \frac{\Sigma}{\rho_{w} \Mach^{2}}
\end{equation}
\noindent
where $\Mach$ is the Mach number of the shock.

In this picture, the density structure in the post-shock region provides the
pressure gradient needed to counterbalance the incoming ram pressure of the
pre-shock material. In a steady state converging flow with constant pre- and
post-shock density and velocity, this additional pressure support would be
unnecessary and the shocked region
would simply behave as a two-shock Riemann problem.
The spatial and temporal variations in the turbulent flow that the
shock
moves
through results in the development of a density gradient in the post-shock region.
For an isothermal fluid, the corresponding density profile can be roughly exponential.
Variations in the velocity and density field, and the non-zero pressure support from
the converging flow behind the shock,
can lead to deviations from this exponential
form, but we expect that the general idea holds. Fluids with different equations of
state, or other sources of pressure support like magnetic fields, could display 
other primary post-shock solutions. We will discuss these possibilities in more 
detail in Section \ref{section:discussion}.

%
%
\subsection{Time-Dependent Exponential Waves}
\label{section:time_dependence}

\begin{figure*}[ht]
\begin{center}
\includegraphics[width=7.1in]{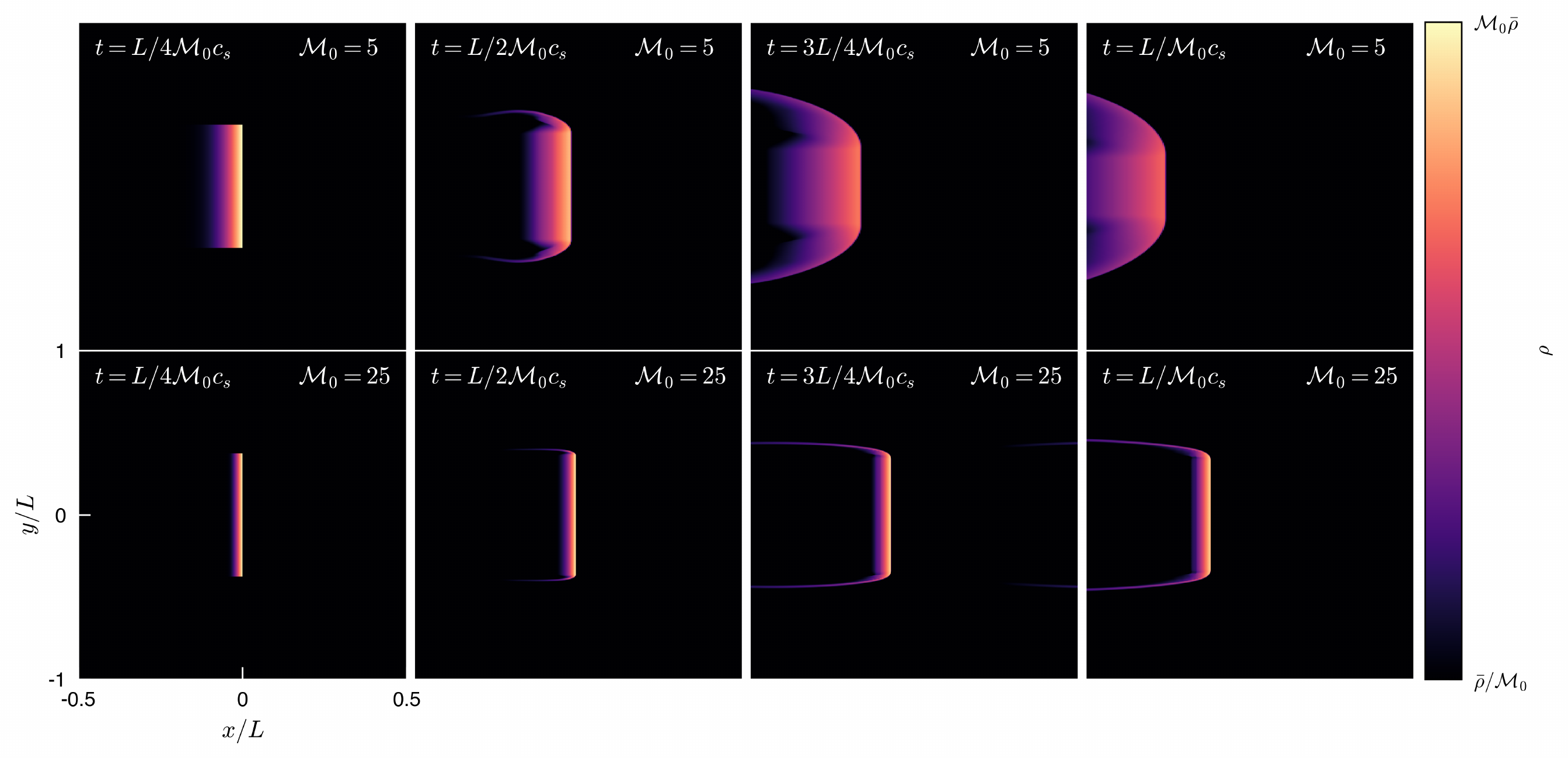}
\caption{\label{fig:exp_waves}
Simulations of exponential waves traveling through a background medium. Shown are thin slices through
the
density distributions for exponential waves initially traveling to the right with Mach numbers of 
$\Mach_{0}=5$ (upper row) and $\Mach_{0}=25$ (lower row), at times $t=[0,1/4\Mach_0,3/4\Mach_0,1/\Mach_0]$ (left to
right). The logarithmic color map spans the range $\rho=[1/\Mach_0,\Mach_0]$.
The initial peak density for
each wave is set to $\rhostar=\Mach_0\rhobar$, while the background medium has density 
$\rho=\rhobar/\Mach_0$. The initial surface density of each wave was set to
$\Sigma_0=0.125\rhobar L$,
with corresponding exponential scale lengths of $h_0=0.025L$
($\Mach_0=5$) and $h_0=0.005L$ ($\Mach_0=25$).
Numerical details of the simulation are discussed in Section \ref{section:idealized_model}.
The simulations demonstrate that the deceleration associated with the ram pressure from the
on-coming pre-shock material causes the shocked regions
to spread behind the shock and
decline in peak density.
}
\end{center}
\end{figure*}

\begin{figure*}[ht]
\begin{center}
\includegraphics[width=7.1in]{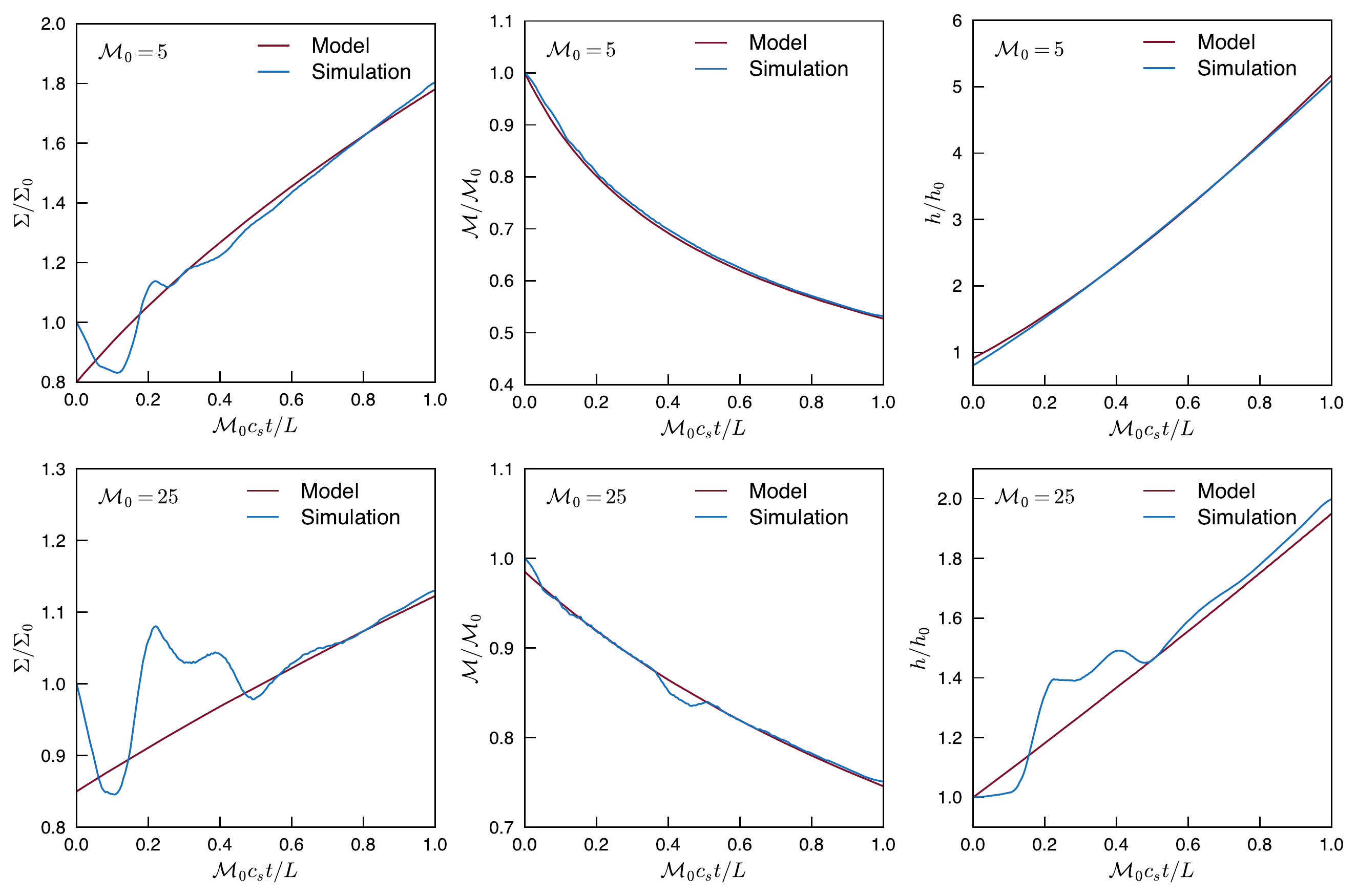}
\caption{\label{fig:wave_model}
Time evolution of properties of exponential waves traveling through a background medium.  
Shown are the surface density (left column), Mach number (middle column), and exponential
scale length (right column) with time of the waves from the 
simulations visualized in Figure \ref{fig:exp_waves}.
The properties of the traveling
waves measured in the simulations are shown as blue lines in each panel, while
the results from the
analytical model described in Section \ref{section:time_dependence} are
shown as red lines. The analytical model for the time dependent surface
density, Mach number, and scale lengths of the exponential waves works well,
and much of the variation owes to uncertainties in separating the
tails of exponential waves 
from the background medium or relaxation from the initial conditions.
}
\end{center}
\end{figure*} 

Motivated by the typical post-shock shape of shocked regions
in our turbulence simulation, 
in Section \ref{section:exponential_atmospheres}
we considered an exponential atmosphere model for 
isothermal shocked regions
traveling through a background medium.
As the 
exponential shocked region
moves through the background medium, 
the region could be decelerated by ram pressure from the
pre-shock material or an increase in its surface density.
The density contrast
between the pre-shock material and the density peak will decline
as the shocked region
decelerates, and as the Mach number of the shock
decreases.
To get some sense of the time-dependence of an isothermal
shocked region
traveling through a background medium, 
we can extend the
exponential atmosphere model to account for the effects associated with
the region's
deceleration.
To do so, we will still approximate the wave as in pressure equilibrium with 
ram pressure from the background. The region will therefore still have an
exponential atmosphere behind the shock,
but the scale length of the atmosphere
will increase as the mass of the wave grows and the wave decelerates.

First, we can model the time-dependent growth of the surface density
of the shocked region.
As the shock
plows through the surrounding background
material, any new material accrued into the shocked region
will depend on the background
density $\rhow$ and the velocity $\vel=\Mach \cs$ of the shock.
With the
ansatz that this material is deposited with some efficiency $\epsilon$, we
can write time rate of change of the surface density as
\begin{equation}
\frac{d\Sigma}{dt} = \epsilon \rhow \Mach \cs.
\end{equation}
\noindent
The value of $\epsilon\ne1$ in general, as not all of the background material
that encounters the shocked region
will become permanently entrained.
The Mach number of the shock 
will change with time, but we can still implicitly
calculate the time-dependent surface density as
\begin{equation}
\label{eqn:surface_density}
\Sigma(t) = \Sigma(t=0) + \int \epsilon \rhow \Mach \cs dt.
\end{equation}
\noindent
The surface density of the shocked region
increases according to the time-integral
of the surface density flux of pre-shock material the shock
encounters, 
moderated by some efficiency parameter $\epsilon$.

The deposition of this material will be accompanied by the deposition of relative
momentum into the shocked region,
and in the case where the background medium is uniform
in density and momentum we approximate 
the rate of this momentum deposition as proportional
to the relative velocity of the background medium with respect to the 
shock
times
the mass accretion rate into the shocked region.
This momentum deposition will reduce the
relative velocity of the shock
and background.  We can balance the rate at which
the relative momentum from the background medium is added to the
shocked region
and the
corresponding rate at which the shock
decelerates.  We can then write
\begin{equation}
\label{eqn:dmdt}
\Sigma \cs \frac{d\Mach}{dt} = -\eta \cs \Mach \frac{d\Sigma}{dt}
\end{equation}
\noindent
where $\eta$ describes the efficiency of depositing momentum from the
background material into the traveling shocked region.
Again, $\eta\ne1$ in general
and the efficiency of mass and momentum deposition do not have to be equal (i.e.,
we have no clear reason to require $\epsilon=\eta$).
For constant mass and momentum deposition efficiencies,
the solution to Equation \ref{eqn:dmdt} is a power-law relation between the Mach number
and the surface density,
\begin{equation}
\label{eqn:mach_number}
\Mach(t) = \Mach(t=0) \left( \frac{\Sigma}{\Sigma(t=0)}\right)^{-\eta}.
\end{equation}
\noindent
If we assume that the density distribution behind the shock
maintains
instantaneous hydrostatic equilibrium, then the pressure gradient behind
the shock
will be balanced by the deceleration from the instantaneous 
ram pressure of the
background material.  We are making the same assumptions that lead to Equations
\ref{eqn:hydro_equil} and \ref{eqn:scale_length} above, but now allow
for the surface density of the wave to change with time according to Equation
\ref{eqn:surface_density}.
The time-dependent scale length can then be modeled as
\begin{equation}
\label{eqn:scale_length_t}
h(t) = \frac{\Sigma(t)}{\rhow \Mach^{2}(t)}.
\end{equation}
\noindent
As the surface density of the shocked region
increases and the Mach number of the
shock decreases,
the scale length of the post-shock density distribution
increases. The
material of associated with the shock
spreads through the post-shock
region. The isothermal jump conditions between the peak density $\rho_0$ 
and the pre-shock
density $\rhow$ are maintained, since $\Sigma = \rho_0 h$ for an
exponential density profile. 

\subsection{Idealized Simulations of Exponential Waves}
\label{section:idealized_model}

Testing the above model of shocked regions
in the context of the turbulence simulations is
difficult because of the complexities of the turbulent flow. Each
shocked region
encounters differing, time-dependent pre-shock conditions and variations
in their locally-convergent velocity field. Instead, we have tried to
test the model for shocked regions via controlled simulations of the motion of exponential waves through
a background medium.  To do this, we use the hydrodynamics code
{\it Athena} \citep{stone2008a} to model an exponential wave with 
initial scale length $h_0$ and surface density $\Sigma_0$ traveling through
a background medium $\rhow$ with an initial relative velocity $\vel = \Mach_0 \cs$.  The
fluid is treated as isothermal with a sound speed $\cs = 1$. We perform
two such simulations, with $\Mach_0=5$ and $\Mach_0=25$.  In both cases, in terms of
a characteristic density $\rhobar = 1$ we set
$\Sigma_0/h_0 = \Mach_0 \rhobar$ and $\rhow = \rhobar /\Mach_0$.  In terms of a 
characteristic scale $L=1$, for the $\Mach_0=5$
simulation we set the initial exponential scale length to be $h_0=0.025L$.  
For the $\Mach_0=25$ simulation, we set $h_0=0.005L$.
The simulations are performed on a three dimensional grid of size $1024\times512\times512$
with periodic boundary conditions. For the $\Mach_0=5$ simulation, the spatial resolution of
the simulation is set by the cubic cell size $\Delta x=1/256$.  For the $\Mach_0=25$ simulation, we
use a resolution of $\Delta x=1/1024$ along the shocked region
and $\Delta y= \Delta z =1/256$ in the plane of the shock.
The exponential shocked regions
are initialized as cylinders oriented along the $x$-axis with
a diameter of $0.75L$ in the $y$-$z$ plane. We evolve each system until the wave interacts with
its own wake after it transverses the volume.

Figure \ref{fig:exp_waves} shows the logarithmically scaled map of the density in the $x$-$y$
plane for the $\Mach_0=5$ (upper row) and $\Mach_0=25$ (lower row) simulations, plotted at
times $t=0$ (far left column), $t = 1/4\Mach_0$ (inner left), $t=3/4\Mach_0$ (inner right), and
$t=1/\Mach_0$ (far right).  The exponential waves are traveling to the right at an initial 
velocity of $\vel(t=0) = \Mach_0 \cs$ relative to the background medium.  The
image frames travel at a constant velocity of $\vel = \Mach_0 \cs$, initially centered on the shock
front.  The apparent motion of the shocked regions
from right to left reflect their deceleration
relative to the background medium (which is traveling through the image frames from right
to left with constant relative velocity $\vel = -\Mach_0\cs$).  In addition to the deceleration,
the decrease in the peak density of the shocked regions
and the increase in the post-shock
exponential
scale lengths are apparent from the density distribution.  
The density distributions of both shocked regions
remain close to exponential for the duration of
the simulations.
At the
edges of the exponential waves bow-like shocks
develop
\citep{vishniac1994a}, and the relative size appears larger for the slower
shocked region
since the vertical spreading of the fluid is limited by the sound speed and the
absolute time scale in the $\Mach_0=5$ simulation is prolonged relative to the $\Mach_0=25$ simulation in 
the bottom panels.

The qualitative evolution of the exponential shocked regions
shown in Figure \ref{fig:exp_waves}
can be quantified from the simulations and compared with the model presented in Section 
\ref{section:time_dependence}.  We estimate the maximum density $\rhomax(t)$ and exponential
scale lengths $h(t)$ of the post-shock regions,
and their velocity relative to the background medium, 
at 100 time steps evenly spaced over the time span $t=[0,1/\Mach_0]$.  The time-dependent
surface densities of the shocked regions
are estimated as $\Sigma(t) = \rhomax h$.  Figure \ref{fig:wave_model}
shows the surface density (left panels), Mach number (center panels), and exponential
scale lengths (right panels) estimated for the shocked regions
in the $M_0=5$ (upper row) and $M_0=25$
(lower row) simulations, normalized to their initial values.  These quantities estimated from
the simulation data are shown as blue lines.  We then use Equations \ref{eqn:surface_density}-\ref{eqn:scale_length_t}
as fitting functions to
model the time-dependence of the simulation data (red lines). The mass
accretion efficiency is taken as $\epsilon=0.88$, while we use momentum
efficiencies of $\eta\approx0.79$ for $\Mach_0=5$ and $\eta\approx1.0$
for $\Mach_0=25$. Relative to Equation \ref{eqn:scale_length_t}, we
allow
for a mildly nonlinear time dependence in the scale length of $\bar{h}(t) \propto h(t)^{\alpha}$ with $\alpha=0.9$ for $\Mach_0=5$ and $\alpha=0.8$ for $\Mach_0=25$.
The early variation apparent in the surface density and scale lengths owes to relaxation from the approximate initial conditions that model the entire
exponential waves as traveling with the same initial group velocity, and in 
inaccuracies in separating the wave from the background medium. These lead to
$\sim10\%$ uncertainties in the measured shocked region
surface density and scale lengths,
and we account for these errors when computing the
time-dependent models shown in Figure \ref{fig:wave_model}.

As Figure \ref{fig:wave_model}
demonstrates, the model presented in Section \ref{section:time_dependence} roughly
recovers the time-dependence of the surface density, velocity, and scale lengths of the
exponential shocked regions
as they are decelerated by the background medium. Physically, this
model succeeds because
the exponential atmosphere behind the shock
responds quickly to the changing ram pressure
from the medium ahead of the shock.

%
%
\section{Shocked Region Lifetimes}
\label{section:shock_lifetimes}

\begin{figure}[t]
\begin{center}
\includegraphics[width=3.3in]{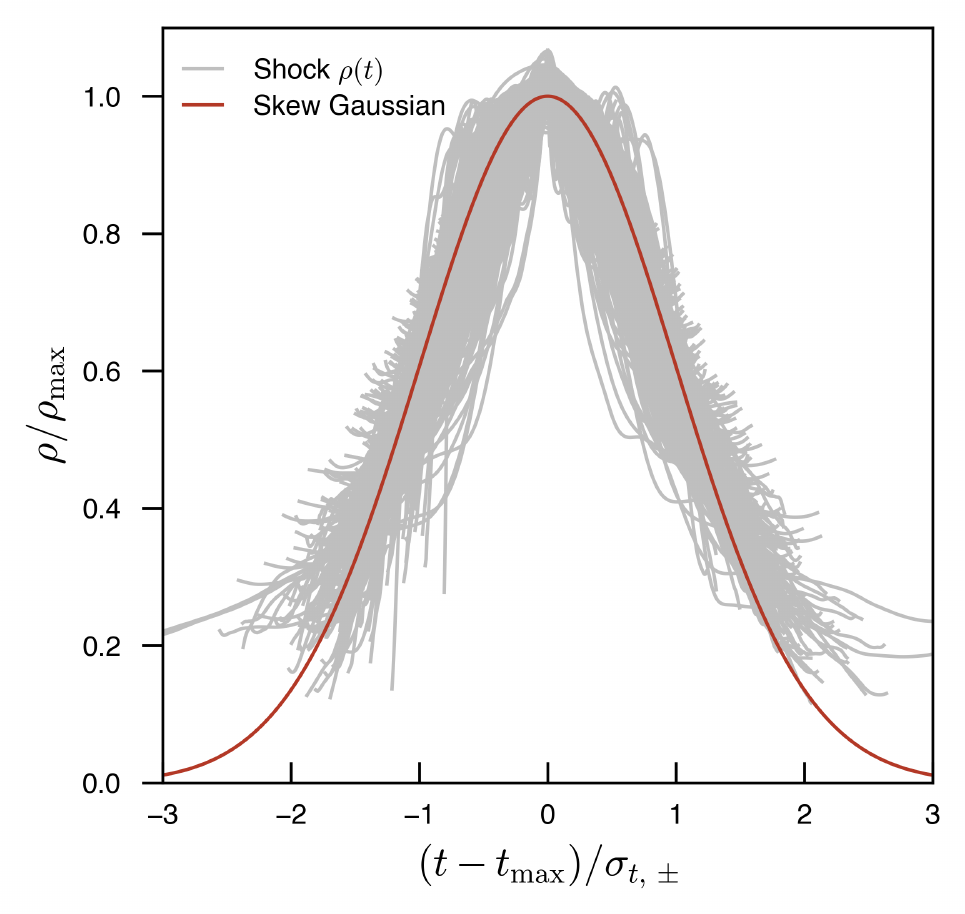}
\caption{\label{fig:density_vs_time}
Time-dependence of the peak density of shocked regions
in isothermal
supersonic turbulence. Shown are the rise and fall of the
density in very dense shocked regions
($\rho>25\rhobar$ at their peak)
in our turbulence simulation (gray lines), measured in between
drivings. The ordinate and abscissa are rescaled by the peak
density and by their rise and fall times determined by a skew Gaussian
fit (red line, see Equation \ref{eqn:skew_gaussian}). Dense
shocked regions
in supersonic isothermal turbulence spread very quickly
and rapidly decline in maximum density.
}
\end{center}
\end{figure}

The comparison presented in Section \ref{section:time_dependence} between
the simulated exponential waves and the time-dependent exponential atmosphere
model suggests that the deceleration of shocked regions
in supersonic isothermal turbulence will lead to a 
rapid decline of their peak densities with time. Using the method for tracking shocked regions
described in Appendix \ref{section:group_tracking}, the time dependence of the peak density of simulated
shocked regions
can be measured.

During the turbulence simulation described in Section \ref{section:simulation} the 
simulation output frequency is increased dramatically after twenty-five turbulent
crossing times, such that five hundred outputs are recorded in between applications of
the driving field. We identify dense regions from tracers with interpolated densities
$\rho\geq 25\rhobar \approx \Mbar^2\rhobar$
in these simulation outputs according to
the method described in Appendix \ref{section:group_finding}. Using the population
of dense regions identified half-way through this high output-frequency period of
the simulation, we track the shocked regions
forward and backward with time following the
method described in Appendix \ref{section:group_tracking}. We then have time
trajectories of each shocked region's
properties over a short period where the simulation
output frequency enables us to follow them reliably. We identify the time at
which each shocked region
reaches its maximum density over this window, and can then
analyze the formation and dispersal of the shocked regions
with time.

As anticipated from the model presented in Section \ref{section:time_dependence},
the individual shocked regions
with high peak densities evolve very quickly. Figure \ref{fig:density_vs_time} shows the rise and fall of shocked regions
with peak densities $\rho_0 \geq 25\rhobar$ (gray lines).
For each shocked region,
we fit a skew Gaussian of the form
\begin{equation}
\label{eqn:skew_gaussian}
\rho(t) = \rho_{\mathrm{max}} \exp\left[ - \frac{(t-t_\mathrm{max})^2}{2 \sigma_{t, \pm}^2} \right],
\end{equation}
\noindent
where $t_{\max}$ is the time of maximum density $\rho_{\mathrm{max}}$. The quantity
$\sigma_{t,\pm}$ equals the rise time $\sigma_{t,-}$ when $t-t_{\mathrm{max}}<0$ and
the fall time $\sigma_{t,+}$ when $t-t_{\mathrm{max}}>0$. The rise and fall times are
fit separately.  We find typical fall times of $\sigma_{t,+} \approx 2-4\times10^{-3} L/\cs$.
The distribution of fall to rise times has a mean of $\sigma_{t,+}/\sigma_{t,-} \approx 1.5$, with 
a tail extending to $\sigma_{t,+}/\sigma_{t,-} \approx 5$.

The lifetimes of dense shocked regions
in the supersonic isothermal turbulence simulation are
quite short, comparable to the sound crossing time across the thickness of the post-shock
region. The portion of their existence when their density is rising is very short, comparable
to the sound crossing time across the cells needed to resolve the density discontinuity at the
shock
interface with the pre-shock material. The time over which their density declines is
only slightly more extended, but substantially longer than the time material takes to flow
from the pre- to post-shock regions. 
The short lifetimes of these shocked regions
may bear on
models of gravitational collapse in turbulent fluids, and we revisit this measurement in that
context in Section \ref{section:gravity} below.

\section{Deceleration and Spreading of Shocked Regions}
\label{section:spreading}

The previous sections have outlined an exponential atmosphere
model for shocked regions, where the
exponential scale length adjusts to the deceleration of the 
shocked region owing to the on-coming ram pressure of the
pre-shock material. This deceleration causes the peak density
of shocked regions to decline as the exponential atmosphere
spreads behind the traveling shock.

The idealized simulations of exponential waves presented in 
Section \ref{section:idealized_model} illustrate this deceleration
and spreading of shocked regions, but demonstrating this effect
for shocked regions in supersonic turbulence requires more
effort. To this end, we have selected a shocked region tracked over the
high-frequency output portion of the $N=512^{3}$ simulation and
measured $\sim2,200$ individual trajectories of the subset of
tracer particles continuously
associated with the shocked region. As the shocked region travels
and spreads, material in the exponential atmosphere is decelerated
and gradually lags behind the shock. Relative to the density peak just
behind the shock, each parcel of material in the exponential atmosphere will
reside at a time-dependent distance $x$ behind the peak. As the material
spreads, the distance $x$ of each parcel will typically 
increase from an initial separation $x_0$ to a larger distance after some
time $t$.

\begin{figure}[t]
\begin{center}
\includegraphics[width=3.3in]{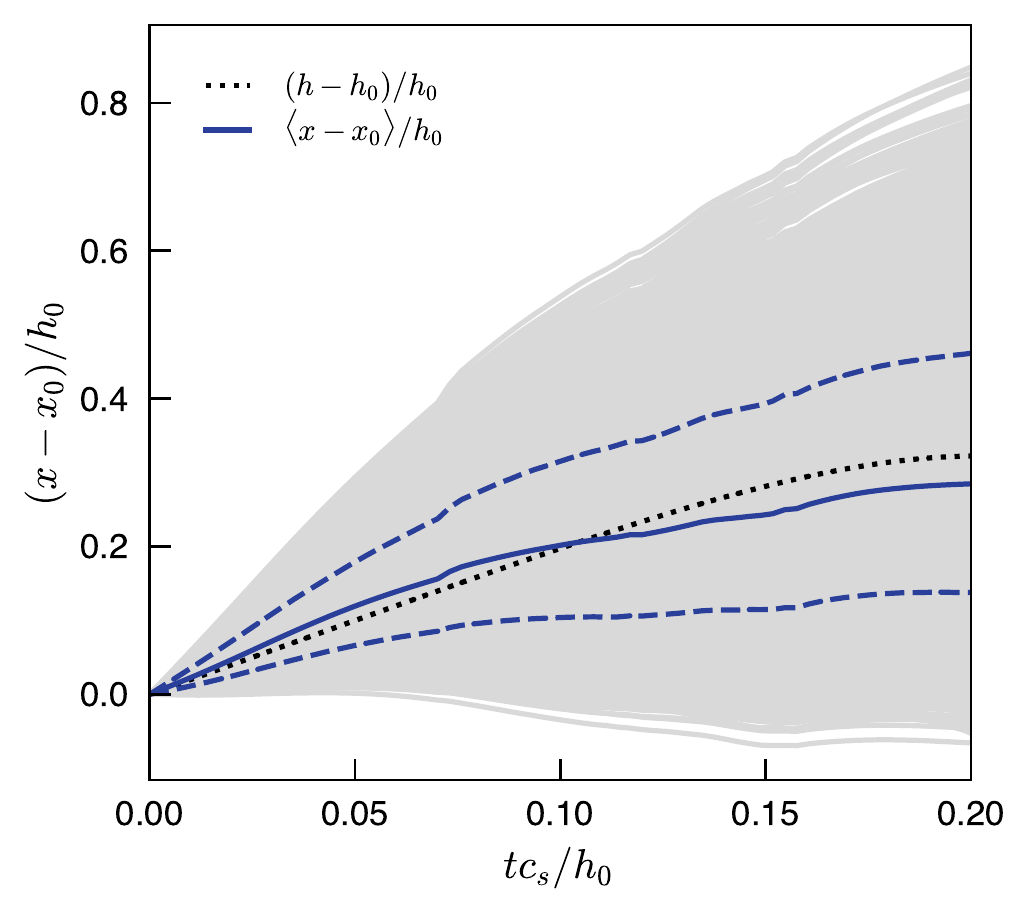}
\caption{\label{fig:spreading}
Spreading of post-shock material in a shocked region with time.
As shocked regions decelerate owing to on-coming ram pressure from
the pre-shock material, their exponential atmospheres spread through
the post-shock region. The gray lines show the time-dependent separation $x$
between $\sim2,200$
tracer particles identified in a shocked region and the location
of the region's peak density, as measured along the direction defined by the
associated shock's trajectory. 
Plotted are the post-shock locations $(x-x_0)/h_0$ relative
to the initial separation $x_0$ from the peak, in units of the initial 
exponential scale length $h_0$
of the region, as a function of time $t$ in units of the sound crossing time
of the initial region scale length $h_0/c_s$.
The blue solid line indicates the average time-dependent
separation, while the dashed blue lines indicate the inner 68\% spread in
separations. Given the deceleration of the region, the expected growth of the
scale length $(h-h_0)/h_0$ 
and the mean separation of material from the density peak
can be estimated from the time-dependent surface density and
on-coming ram pressure (dotted black line). The rate of
spreading of material in the
post-shock region is consistent with expectations computed from the deceleration
of the region.}
\end{center}
\end{figure}

Figure \ref{fig:spreading} shows the time-dependent separation
between the tracked tracer particles and the moving density peak, $(x-x_0)/h_0$,
in units of the initial scale length $h_0$ of the atmosphere (gray lines).
In Figure \ref{fig:spreading},
the coordinate $x$ corresponds to the direction of travel of the shock
and increases in the post-shock direction, and time $t$ is scaled by
the sound crossing time across the initial scale length $h_0$.
The tracers spread at a range of rates as they respond to the deceleration
of the shocked region, 
which owes both to their initial distribution of $x_0$ throughout
the post-shock flow and to the interpolation scheme
used to compute the particle velocities. The mean separation of the
tracers and the moving peak of the shocked region is shown as a solid blue
line, and the inner 68\% of the distribution of separations is shown with
dashed blue lines.

To verify that the region spreads in response to the deceleration of the
shocked region, we must estimate the expected rate of spreading.
For an exponential atmosphere extending to zero density,
the density-weighted average distance of material from the
peak is equal to the scale length $h$. For finite atmosphere the mean distance
from the peak is less than $h$, and for this region that extends for $\sim 2 h$
before reaching the background density the mean distance is $\sim0.7h$. The
rate of change of the mean distance from the peak should be proportional to
$dh/dt$.
If the
scale length of the region $h\sim \Sigma/\rho_0$, where $\Sigma$ is the
surface density of the region and $\rho_0$ is the peak density, then
we can write
\begin{equation}
\label{eqn:spreading}
\frac{dh}{dt} = \frac{1}{\rho_0}\frac{d\Sigma}{dt} - \frac{\Sigma}{\rho_0^2}\frac{d\rho_0}{dt} = \frac{1}{\rho_0}\frac{d\Sigma}{dt} - \frac{\Sigma}{\rho_0}\frac{d\log \rho_0}{dt}.
\end{equation}
\noindent
We can integrate this equation to find the expected spreading of the region
$(h-h_0)/h_0$
relative to the initial scale length $h_0$ (dotted black line). Equation \ref{eqn:spreading} accounts for how changes in the surface density of the region
affect its deceleration in addition to the response to the on-coming
ram pressure.

As Figure \ref{fig:spreading} demonstrates,
we find reasonable agreement between the spreading rate
measured
from the tracer particle positions and the estimate computed from the
expected time-dependence of the scale length. We present this measurement
as supporting evidence that the exponential atmosphere model provides a
useful description of the post-shock flows in shocked regions. We caution
that this smooth behavior of spreading only occurs when a region travels
through a pre-shock region with relatively constant density and velocity.
If instead the region is further compressed to form a higher density sheet, the
spreading will cease at least momentarily. Nonetheless, when the pre-shock
conditions allow for the exponential atmosphere to develop it spreads at
a rate comparable to expectations based on the region's deceleration.

%
%
\section{Properties of Shock Populations}
\label{section:shock_populations}

The preceding analysis has explored the properties of individual 
shocked regions
and the average shocked region
properties determined from the population of
dense regions identified in the simulation volume. We now turn to the
properties of the population of shocked regions
as a whole. In analogy with 
treating dense regions of a cosmological density field in the context
of a ``halo model'', we will measure some important properties of
the dense regions of turbulence in the context of a ``shock model''
for the population of shocks and associated shocked regions
present in the fluid.
Dense regions of the turbulent fluid are assigned to individual
shocks
identified using the method described in Appendix \ref{section:group_finding}.
The locations of shocked regions are taken to coincide with the
their density maxima. Their interior density profiles are
assumed to follow the interpolated density at the locations of
the tracer particles assigned to them.

%
%
\subsection{Spatial Clustering of Dense Regions}
\label{section:clustering}

\begin{figure}[t]
\begin{center}
\includegraphics[width=3.3in]{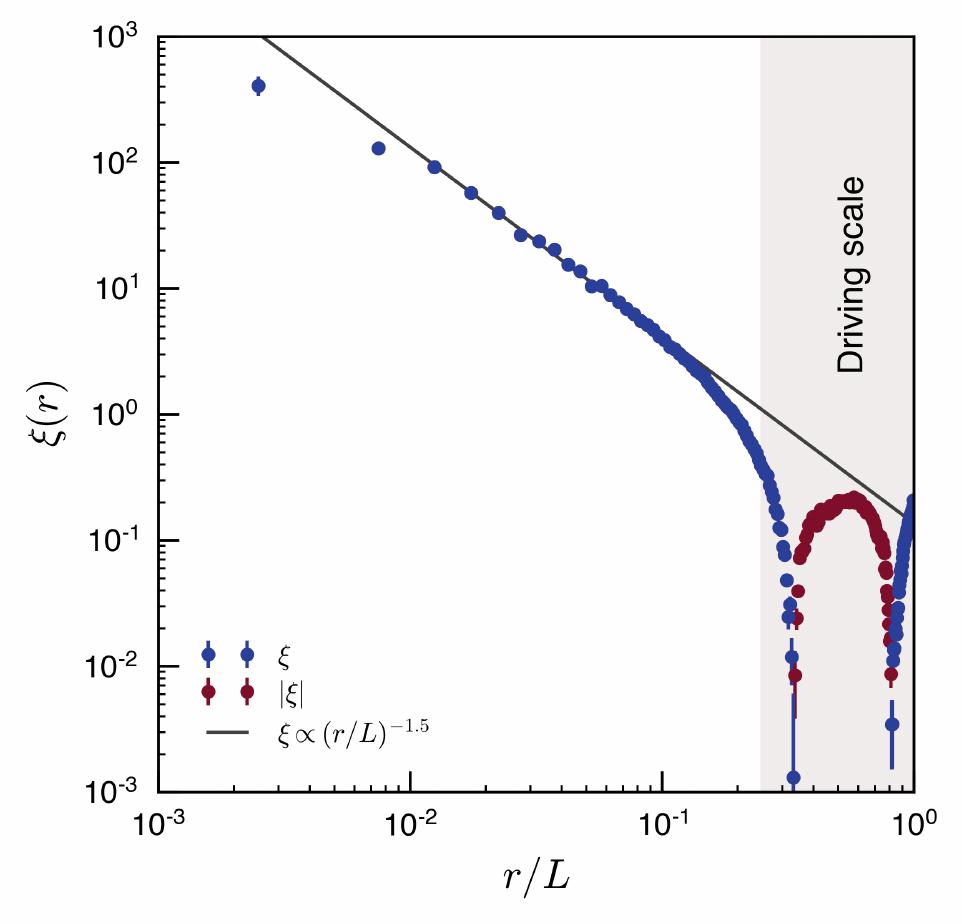}
\caption{\label{fig:correlation_function}
Two-point correlation function of dense shocked regions
in supersonic
isothermal turbulence. Shown is the correlation function 
of the density maxima of shocked regions
with peak densities $\rho_0\gtrsim\bar{\Mach}^2\rhobar$,
computed using the \citet{landy1993a} estimator, constructed by
counting pairs of dense regions
in bins of radial separation and comparing them
with the corresponding clustering of randomly distributed locations. The
correlation function is measured for five statistically-independent times
during the turbulence simulation, and then averaged. The error bars indicate
the relative uncertainty per radial bin, scaling with the square root
of the number of pairs in each. On small scales, the shocked regions
are strongly clustered
(blue points), with the correlation function scaling approximately 
as $\xi\propto(r/L)^{-1.5}$ (black line). At scales comparable to the driving
scale (gray shaded region), the dense regions
become anti-correlated with $\xi<0$ ($|\xi|$ is shown
with maroon points).
}
\end{center}
\end{figure} 

The density field shown in Figure \ref{fig:box} illustrates some important
features of shocked region
population in isothermal turbulence. First, the densest regions are
spatially clustered. In projection, the sheet-like structures associated with
shocks
appear filamentary. For turbulence driven at low spatial frequencies,
the densest shocked regions
occur near the intersections of
large-scale velocity perturbations. Second, there are large interior regions in the
turbulent fluid, comparable
to the driving scale, that are nearly devoid of dense shocked regions.
As discussed in
Section \ref{section:dense_regions}, these regions have volume filling densities
$\rhobar/\Mbar \lesssim \rho \lesssim \rhobar$ and represent mild rarefactions owing to the
large-scale driving modes. Once the shocked regions
are identified using the method
described in Appendix \ref{section:group_finding}, the statistical properties
of their spatial distribution should reflect these features.

A useful statistic familiar from cosmology is the two-point correlation function
$\xi(r)$ that describes the excess probability for two points pulled from
a spatial distribution to be separated by a distance $r$ relative to two points
pulled from a uniform random distribution. A convenient estimator for $\xi(r)$
was provided by \citet{landy1993a} in the context of galaxy surveys, written as
\begin{equation}
\label{eqn:landy-szalay}
\xi(r) \approx \frac{(DD - 2DR + RR)}{RR},
\end{equation}
\noindent
where $D$ represents the locations of data points for which $\xi(R)$ is desired,
and $R$ represents the locations of randomly distributed points with the same
average number density. The quantities $DD$, $DR$, and $RR$ correspond to 
data-data, data-random, and random-random point pairs separated by a distance
$r$.

To compute $\xi(r)$ for shocked regions
with peak densities $\rho_0\gtrsim\Mbar^2\rhobar$ 
in our turbulence simulation, we identify the locations of maximum density for each
region to generate our data sample $D$. We then generate a uniform random
point distribution of the same size to populate $R$. The point populations are
loaded into k-D trees, allowing for fast neighbor searching to find point pairs
separated by a distance $r$. The $DD$, $DR$, and $RR$ pairs are computed, and Equation
\ref{eqn:landy-szalay} used to estimate $\xi(r)$. To build signal and to average over
high time-frequency variations in the correlation function, the process is repeated for
five statistically independent times during the simulation and the measurements averaged.
Figure \ref{fig:correlation_function}
shows the resulting two-point correlation function for dense regions
in the
turbulence simulation. The correlation increases to small scales, behaving
as a rough power law with $\xi(r)\propto (r/L)^{-1.5}$. On scales of a few cells,
the correlation weakens somewhat, but this slight turn-down may owe to our
shocked region
identification method or to resolution effects near the grid scale. On
scales comparable to the driving scale of the turbulence the dense regions
become
anti-correlated, reflecting the presence of large, underdense voids in the
density distribution.

Further analogies with the spatial correlations of dark matter halos may provide
additional insight. The densest shocked regions
in supersonic turbulence
are clearly more strongly clustered than
the density field, which is itself spatially clustered. The analog in cosmology
is the concept of halo bias, where the ratio of the halo and matter correlation
functions is $b>1$ for strongly-clustered dark matter halos. The base analytical
picture for understanding halo bias is the peak-background split model \citep[e.g.,][]{mo1996a,sheth1999a,tinker2010a},
where the excess abundance of halos in regions of enhanced background density 
can be used to estimate their clustering bias relative to the matter field.
For turbulence it may be tempting to imagine a
``peak density function''
$dn/d\rho_0$ describing the differential number density of shocked regions
as a function
of their peak density, or a mass function in analogy to the halo mass function $dn/dM$,
which then could be used to estimate the expected
bias relative to the turbulent density field. Indeed, similar
ideas have been explored before in the context of driven and decaying
turbulence \citep[e.g.,][]{smith2000a,smith2000b}.
The excursion set formalism
model of \citet{hopkins2013a} can be used to compute an analytical model for
the clustering of dense regions in turbulence \citep{hopkins2013b} and predicts
that $\xi(r)\propto r^{-1}$ on small scales, steepening to $\xi(r)\propto r^{-2}$
on large scales. Our findings appear roughly
consistent with these predictions,
but our group finding algorithm, which prevents the identification of distinct
regions near the grid scale, does not enable us to confirm robustly the origin of
the flattening of $\xi(r)$ on small scales.
We leave additional comparisons for future work.

%
%
\subsection{Dense Regions and the Density PDF}
\label{section:dense_pdf}

\begin{figure}[t]
\begin{center}
\includegraphics[width=3.5in]{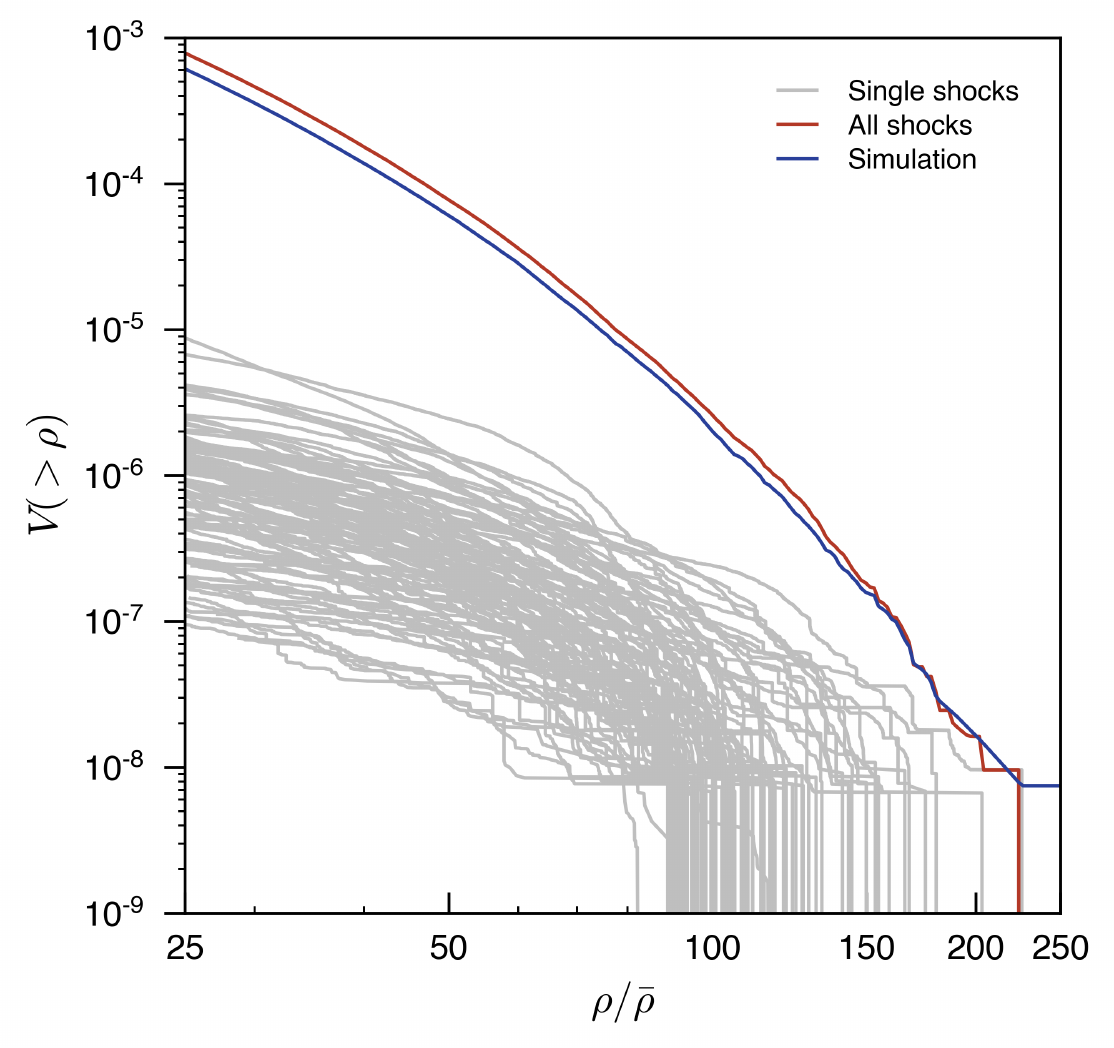}
\caption{\label{fig:density_cdf}
Cumulative density distribution function for supersonic isothermal turbulence,
reconstructed from the combined density distributions of individual
shocked regions.
The volumetric density probability distribution function of supersonic isothermal
turbulence is close to lognormal (e.g., Equation \ref{eqn:pdf}), and the cumulative density distribution function in turbulence corresponds to the fraction of volume in the fluid above
a given density. The density CDF measured from the turbulence simulation (blue) can be compared
with the density CDF constructed by summing the individual density distributions (gray lines) of all of the dense regions identified using tracer particles in the simulation (red line).
The slight excess of the reconstructed CDF over the simulation CDF results from the allowed
 overlap of the Voronoi tessellations used to estimate the density PDFs of individual shocked regions.
 Overall, the excellent agreement demonstrates that the dense tail of the 
 turbulent density PDF corresponds to a population of distinct structures.
}
\end{center}
\end{figure} 

The origin and shape of the density probability distribution function (PDF)
of supersonic isothermal turbulence influences the star formation process.
The roughly lognormal shape of the PDF has been cited as evidence for a
statistical origin \citep[e.g.,][]{vazquez-semadeni1994a,padoan2002a}.
However the character of the forcing field influences the shape of the PDF
on the high-density tail \citep[e.g.,][]{federrath2008a,federrath2010a,hopkins2013c}, 
with compressive
modes leading to more high-density material. This result
implies that the statistics of the turbulence at high densities retains
some memory of the properties of large-scale driving modes, which may
argue against the density PDF arising simply from central limit theorem
statistical arguments.
In the context of this work, where we have identified individual
dense regions in supersonic turbulence using the method described in Appendix \ref{section:group_finding} and tracked their time evolution following Appendix \ref{section:group_tracking},
a clear test for our model of dense regions in supersonic turbulence as a collection of distinct traveling waves is whether the density PDF can be reconstructed from their properties.

The turbulence simulation performed on a Cartesian mesh described in Section \ref{section:simulation} does display a nearly lognormal density PDF with a width appropriate for its root-mean-squared Mach number. For convenience, we will work with the density cumulative distribution
function (CDF)  
\begin{equation}
\label{eqn:cdf}
V\left(>\rho\right) = \int_{\rho}^{\infty} \frac{dp}{d\rho} d\rho,
\end{equation}
\noindent
where the density PDF $dp/d\rho$ is normalized to integrate to unity over all densities $\rho>0$.
Figure \ref{fig:density_cdf} shows the density CDF for our turbulence simulation (blue line), 
computed by summing the volume in cells above a given density. We examine only the tail
of the PDF at densities $\rho>25\rhobar$, corresponding to the threshold density for the
tracer particles we associate into groups.

Recovering the density CDF from the individual dense regions in our catalogue constructed
from tracer particles is more involved. By design, the density field interpolated at the tracer
particle locations varies on scales less that the cell width $\Delta x$. To
compute the density CDF from the tracer particles therefore requires us to assign a
volume to each tracer particle, and then sum the volume occupied by tracers in our catalogue
above a given density. For each group in our catalog we use the {\it Voro++} library 
\citep{rycroft2009a} to construct
a Voronoi tessellation about the tracer particle positions, accounting for the presence of 
nearby low-density tracers that surround each group. Individual tracer particle 
groups then have their own density CDF $V_{i}(\rho>)$, shown as gray lines in 
Figure \ref{fig:density_cdf} for the one hundred identified groups with the highest
peak densities. The total density CDF of the tracer particle groups then corresponds to the
sum of the individual group CDFs, e.g., $V(\rho>) = \sum_i V_{i}(\rho>)$. The resulting
total density CDF reconstructed from our group catalogue is shown as a red line in Figure
\ref{fig:density_cdf}. The agreement between the simulation and reconstructed density CDFs appears quite good, and this result is nontrivial. The slight excess in the reconstructed
CDF owes to a combination of our interpolation scheme (here, PPI is used) and permitting
overlap of the volumes assigned to the individual groups (tessellating about all tracer
particles in the simulation simultaneously would avoid this). Note that in general interpolation
schemes that smooth the density field near maxima will not lead to an accurate reconstructed
density CDF, as the highest density tail of the CDF will be suppressed.

This demonstrated correspondence between the simulated and reconstructed density CDFs demonstrates that this statistical property of the turbulence arises from the
internal structure of distinct regions \citep[for some related analytical models, see][]{fischera2014a,fischera2014b,myers2015a,veltchev2016a,donkov2017a}.
Projections of multiple physically distinct regions along the line-of-sight will then comprise
the filaments that produce the observed column density PDF \citep{moeckel2015a,chen2017a}.
Each of the individual regions shown in Figure \ref{fig:density_cdf} have been tracked with time during a portion of the simulation, and we
have checked that time variations in the simulation density CDF correspond to the
evolution of the density structures of individual groups as described in
Section \ref{section:shock_lifetimes}. 
We can confirm that the rapid evolution in the peak density of
the densest regions discussed in Section \ref{section:shock_lifetimes} indeed corresponds to
the time variation in the high-density tail of the PDF (and CDF), as suggested
by the models of \citet{hopkins2013c}.
Indeed, conceptually the reconstructed density CDF
can be considered as an integral over a
peak density function $dn/d\rho_0$ times
the internal density PDF of an individual shocked region
with peak density $\rho_0$. However, the 
group-to-group variations apparent in Figures \ref{fig:density_profiles} and 
\ref{fig:density_cdf} may suggest a more complex picture. We speculate that variations
in the density PDF in turbulence with differing driving mechanisms, or with differing
physics (e.g., magnetic fields, adiabatic equations-of-state), will correspond conceptually
to changes in the number density and the
typical density profile of regions with a given peak density. In self-gravitating regions,
this connection appears as the development of a power-law tail in the PDF as the density
profile in collapsing regions \citep{kritsuk2011a,ballesteros-paredes2011a,lee2015a,burkhart2017a,murray2017a}.

%
%
\section{Gravity and the Fates of Dense Regions}
\label{section:gravity}

\begin{figure*}[ht]
\begin{center}
\includegraphics[width=7.1in]{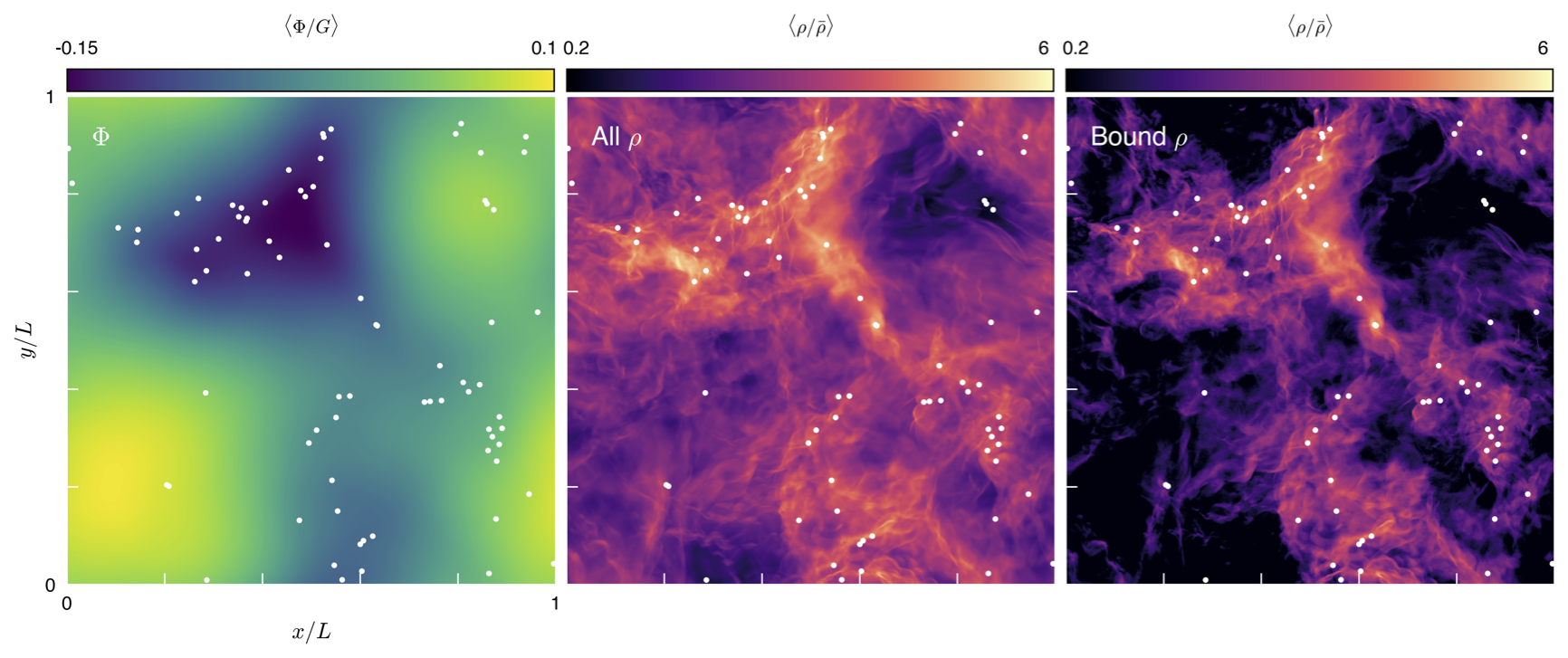}
\caption{\label{fig:group_locations}
Gravitational potential minima in supersonic isothermal turbulence, and their relation to the
density field.
Shown are projections of the gravitational potential (left panel; linear projection) and density field (center panel; logarithmic projection) through the entire
simulation volume shown in Figure \ref{fig:box}. The white points indicate local minima in the gravitational potential, as determined from the potential interpolated at the position
of tracer particles. Density maxima correspond to local potential minima, but the global potential minimum in turbulence does not occur at the maximum density (the densest regions
occupy very small volumes and contain small mass, see Figure \ref{fig:mass_fractions}).
Assuming the turbulence
is marginally self-bound to set the value of the
gravitational constant, regions in the turbulent
fluid that are gravitationally bound to any potential minimum can be identified (right panel; same scale as center panel). Typical gravitationally-bound regions have densities of a few
times the mean density, corresponding to the largest density regions that
contain substantial mass.
}
\end{center}
\end{figure*}

The correspondence between the simulated density distribution of the turbulent fluid and the
density CDF reconstructed from individual regions suggests that the bulk properties of
high-density volumes in turbulence are tightly connected with the detailed properties of distinct
shocked regions.
This picture may therefore have
important ramifications for models of star formation that
involve the turbulent density PDF, such as models that use the density PDF to set the star-formation efficiency of a molecular cloud, the stellar initial mass function, or the core
mass function \citep[e.g.,][]{krumholz2005a,padoan2011a,hennebelle2008a,hennebelle2011a,federrath2012a,hopkins2013a}. 
As our analysis has illustrated, dense
regions are far from static cores and at any given density the density PDF is comprised from differing regions within distinct structures with a range of peak densities.
However, to say much more we need to develop some
expectations for the evolution of the turbulent gas under the influence of self-gravity.

To compute the gravitational potential of the simulated fluid, we solve the Poisson
equation using standard Fourier methods. The Fourier transform of the density
field is computed in the three-dimensional volume using the NVIDIA {\tt cuFFT} library.
The potential is calculated by multiplying the density transform by factors of the wave
number, and then taking the inverse transform. This process provides the potential 
$\Phi$ in
units of the gravitational constant (i.e., $\Phi/G$). We can then interpolate the
potential at the locations of tracer particles to aid in our analysis.

A projection of the resulting gravitational potential is shown in the left panel of
Figure \ref{fig:group_locations}, along with the density field generating the
potential (Figure \ref{fig:group_locations}, center panel). The morphology
of the potential $\Phi$ follows large overdensities in the turbulent fluid, with
broad minima of the potential corresponding to regions with typical density of
a few times the background density. This correspondence results from the
density structure of the turbulence, since regions with $\rhobar \lesssim \rho \lesssim \Mbar\rhobar$
contain a plurality of the fluid mass. 
While density maxima do lie at local
minima in the gravitational potential
(shown as white points in Figure \ref{fig:group_locations}), the global minimum of the potential does not
correspond to a prominent maximum of the density field. The densest regions in the
turbulence carry very little mass, and do not dominate the global structure of the
gravitational potential sourced by the fluid.

To compute gravitationally-bound regions, the value of the gravitational
constant $G$ must be chosen.
To set the value of $G$ we
assume that the gravitational potential energy in the simulation
volume approximately equals the kinetic energy in the 
turbulent motions, such that the entire box is marginally
self-bound. We then have that
\begin{equation}
\label{eqn:virial}
G\rhobar^2 L^5 \approx \frac{1}{2} \rhobar L^3 \Mbar^2 \cs^2,
\end{equation}
\noindent
or solving for G we have
\begin{equation}
G = \frac{\Mbar^2 \cs^2}{2 \rhobar L^2}.
\end{equation}
\noindent
Choosing different geometrical factors of order unity in
Equation \ref{eqn:virial} would not change the results of
our analysis.
With the gravitational constant selected, the tracer particles
associated with any potential minima are identified by using
a friends-of-friends algorithm similar to that described in
Appendix \ref{section:group_finding} with a linking length set
to the cell width $\Delta x$. The relative potential between
the minimum and the maximum potential at the edge of the FOF
groups are computed, and then compared with the relative kinetic
energy of each particle with respect to the potential minima.
Tracer particles with a negative relative total energy are
considered to be bound. This process is analogous to that
commonly performed when identifying substructure in simulated
dark matter halos \citep[e.g.,][]{knebe2011a}, except that
the geometry in turbulence is more complicated.
A logarithmic density projection of the bound regions in the turbulence
simulation is shown in the right panel of Figure \ref{fig:group_locations}.
Most of the mass in bound regions reside in broad potential minima and
at typical densities of a few times the mean. The bound regions
collectively comprise about $50\%$ of the mass of the entire
cloud.

%
%
\subsection{Time Evolution of the Potential Field vs. Dynamical Timescales}
\label{section:potential_evolution}

\begin{figure*}[t]
\begin{center}
\includegraphics[width=7.1in]{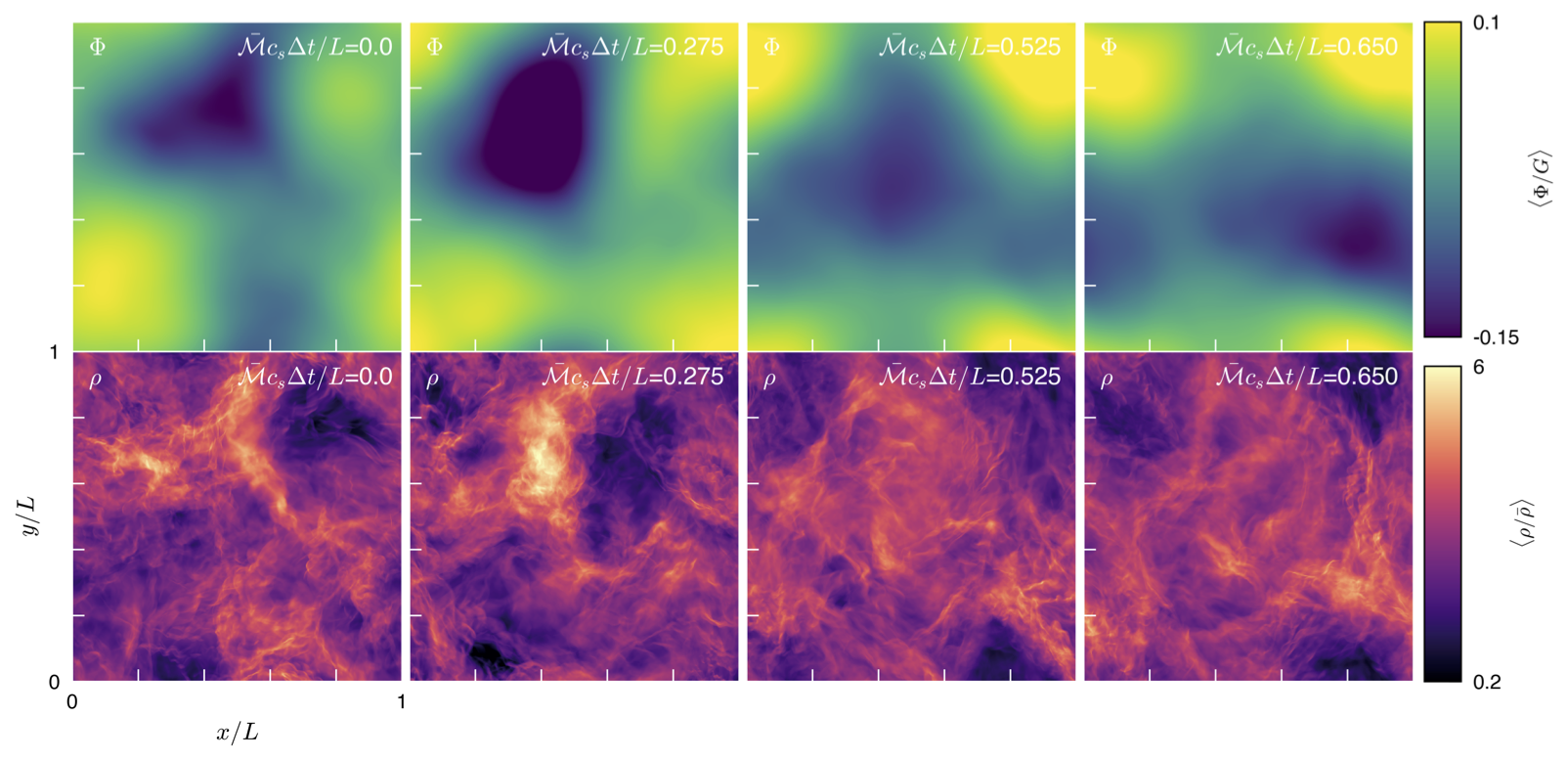}
\caption{\label{fig:potential_evolution}
Time evolution of the large-scale potential field in supersonic
isothermal turbulence. The gravitational potential (top row; linear projection),
sourced by the turbulent density field (bottom row; logarithmic projection),
has broad minima that correspond to regions of moderate density that carry a
significant fraction of the mass of the fluid. The simulation is evolved forward in time (left to right, in units of the mean Mach crossing time across the simulation volume $\tcross \approx L/\Mbar cs$) without the effects of self-gravity to estimate the lifetime of prominent potential minima. Absent
the effects of self-gravity, these potential minima would survive only a few
Mach crossing times of the dense region sourcing the potential minimum. After
this time, the overdensities that source the potential minima disperse (and
others reform).
}
\end{center}
\end{figure*} 

Dense regions in supersonic isothermal turbulence evolve quickly, as discussed in Section \ref{section:shock_lifetimes} above and elsewhere \cite[e.g.,][]{klessen2000a,vazquez-semadeni2005a,glover2007a}. For dense regions to collapse,
their gravitational free-fall time must be shorter than their expansion timescale.
In our turbulence simulations where we identified and tracked individual dense
regions, the expansion time scale is of order the sound crossing time from the pre-
to post-shock regions about the density maxima and is comparable to a few thousandths or one hundredth of the sound crossing time of the whole simulation volume. Dense regions that do collapse and become bound may initially have falling densities or be newly forming
during the collapse of a bound region on larger scales. 
The evolution of the gravitational potential on the scales of the largest bound
regions sets the time scale over which denser interior regions must become bound and
collapse. Since we found in Section \ref{section:gravity} above that large bound regions in turbulence characteristically have
intermediate densities $\rhobar \lesssim \rho \lesssim \Mbar \rhobar$ that contain
substantial mass, we need to monitor the density and potential fields on scales of
the simulation volume over time to determine their typical lifetimes.

Figure \ref{fig:potential_evolution} shows the time evolution of the potential
(upper panels; linear projection) and density (lower panels; logarithmic projection) fields in our simulation volume. The potential field is computed
from the density field following the method described in the previous Section \ref{section:gravity}, but the resulting gravitational acceleration is not
applied to the fluid with the goal of monitoring the typical lifetime of
moderate density regions and their resulting potential minima. Shown are
four separate times during the simulation at $\Delta t = [0, 0.275, 0.525, 0.65]$
times the Mach crossing time $t_{\Mach} = L/\Mbar \cs$. The density enhancement
 in the upper left quadrant at time $t=0$ contains an average interior density
 close to $\rho \sim \Mbar\rhobar$ and sources the broad potential minimum 
apparent in the corresponding image of the potential field. The initial
size of this region is $l \sim L/3$. As the time sequence in
Figure \ref{fig:potential_evolution} shows, the region disperses over a time
$\Delta t \sim 0.5-0.6 L/\Mbar\cs$ that corresponds to $1-2$ Mach crossing
times $\tcross \sim l/\Mbar\cs$ across the region.

The mass-bearing structures with intermediate densities and their
corresponding potential minima have lifetimes that are a few times shorter than the Mach crossing
time across the entire box size $L$ but considerably longer than the typical
lifetime of the densest turbulent structures. In a real molecular cloud
transitioning from low to high mean densities and subsequently undergoing star
formation through the gravitational collapse of its dense interior regions, the
lifetime of the intermediate density regions will influence how star formation
proceeds. For a cloud in rough virial balance, bound regions with
intermediate densities $\rhobar\lesssim\rho\lesssim \Mbar \rhobar$ will have a 
collapse time scale of  $\tcol \lesssim \Mbar^{-1} L/\cs \approx \tcross$. If the
entire mass of such a region were to collapse to form stars, the star formation
efficiency in the entire cloud would be $>10\%$ rather than 
$\sim1\%$ \citep[e.g.,][]{krumholz2007a}. Significant density enhancements within the
intermediate density region will collapse on much faster time scales, and if the
star formation process of these regions supplies the energy that eventually regulates
the cloud's star formation efficiency such mechanisms must therefore operate on time scales
shorter than $\tcol \sim \Mbar^{-1} L/\cs$.

%
%
\section{Discussion}
\label{section:discussion}

This work presents a model where the dense regions of supersonic
isothermal turbulence have density profiles that develop
approximately exponential atmospheres (Sections \ref{section:density_profiles} and \ref{section:exponential_atmospheres}).
We present some idealized simulations of exponential waves and an analytical
model that shows the time-evolution of dense regions may be understood
by accounting for the interaction of the traveling wave with the pre-shock
medium (Section \ref{section:time_dependence}).
Observationally, filaments are seen to display a wide range of profiles that
are generally modeled with power-laws \citep[e.g.,][]{arzoumanian2011a,kirk2013a}. We note that
outside of the smallest scales that may be affected by the beam shape, the
density profiles are not far from exponential.
In simulations, filamentary profiles have been often been
modeled with power-law and Gaussian profiles
\citep{gomez2014a,smith2014a,smith2016a,federrath2016a}. We expect that exponential profiles
will provide comparable quality fits, and benefit from a physical model for their origin. We will examine this issue in future work.
We note
that filamentary profiles in simulations and observations are frequently treated as a radial
profile, whereas our model describes the density profile perpendicular to the shock.
The dense
regions in our simulations are asymmetrical, with much steeper density profiles (e.g., jumps)
ahead of the shock
than in the post-shock region and are oriented mostly along the velocity
field. Symmetrical fitting of the filament profiles do not account for these features.
We note that the power-law radial profile behavior seen in self-gravitating regions of
turbulence arises from physics we do not model \citep{kritsuk2011a,fischera2012a,heitsch2013a,heitsch2013b,federrath2016a,murray2017a,mocz2017a,li2017a}.

\subsection{Caveats to the Exponential Atmosphere Model}

A model that approximately describes the properties of 
dense regions in supersonic turbulence can provide a
useful conceptual picture for the formation and
evolution of dense shocked regions
that may collapse to form
self-bound regions. The applicability of the model
depends on a host of approximations that we have
employed to make the problem tractable, and our
numerical simulations have limitations.
We now examine
some of these assumptions and limitations, and attempt to evaluate, at
least qualitatively, how they might affect the realism
of the physical picture presented in this work.

\subsubsection{Equation of State}

An immediate concern is the applicability of isothermality
to dense regions of observed molecular clouds. The
radiative efficiency of shocks
in dense gas motivates the
isothermal assumption, but the gas does not have to 
remain perfectly isothermal. If the adiabatic index
$\gamma>1$, then the additional pressure support of the
fluid during compression will resist the large amplification
of the density possible in isothermal shocks.
According
to the Rankine-Hugoniot conditions, the factor of $\sim4$
density amplification in individual adiabatic
shocks
may
have a variety of implications to the model. In terms
of the volumetric density PDF, regions of high density
can still develop through (relatively larger numbers of)
successive generations of shocked regions
\citep[][]{scalo1998a,federrath2012a}. But how will
the structure of individual shocked regions
change? The alteration
will depend on whether the shocked regions
remain thin enough
that the time for their interior structure to adjust
to the ram pressure force applied by on-coming material
remains shorter than the time for them to travel an
appreciable distance or their mass to change
significantly. The primary influence on this thickness
will be the value of $\gamma$, but if the adiabatic
index is low enough to allow a fast response to 
ram pressure variations at the shock
front then the
shocked region
structure will reflect the pressure gradient
required under the stiffer equation of state to balance
the ram pressure force.
Indeed, previous simulations of turbulence that include radiative
cooling suggest that the post-shock region will remain close to
isothermal behind a radiative shock front \citep[e.g.,][]{pavlovski2006a}.

\subsubsection{Magnetic Fields}

We do not attempt to generalize
our results to MHD turbulence. Models for the
statistical properties of strong, incompressible
MHD turbulence have 
been developed \citep{goldreich1995a} and examined
in the context of large scale numerical simulations
\citep[][see also \citealt{perez2012a}]{beresnyak2011a,beresnyak2014a}.
The shock compression
of regions threaded by weak magnetic
fields will lead to an amplification of the field and
an increase in the magnetic pressure support within the
fluid, thereby changing the balance between internal 
pressure and exterior ram pressure. The density PDF
of MHD turbulence does display an approximately 
lognormal distribution \citep[e.g.,][]{ostriker2001a}, which
suggests large density inhomogeneities that would allow
for dense shocked regions
to encounter a low density background 
a short time after their formation. However, their pressure
support and internal structure could differ significantly 
from isothermal hydrodynamic shocks.
Analysis of filaments
in MHD simulations show that the central density contrast
of the filaments are reduced, but that the overall profile
of the filaments may not change substantially \citep{federrath2016a}.

\subsubsection{Limitations of the Simulations}

The numerical
simulations of supersonic turbulence we present 
have limitations. 
Given the reconstruction scheme
used, the first scale length of the exponential atmosphere
of the densest shocked regions
($\rho\gtrsim50\rhobar$) will not be well-resolved.
As Figure \ref{fig:density_profiles} indicates, our ability
to study the density profile very near the shocked region
peaks 
(e.g., $x\lesssim0.5h$) will affected by resolution. 
However, the exponential profile appears to be consistent
with the actual measured density profile of shocked regions
out to 
typically $x\sim3h$ where even dense shocked region
atmospheres have been 
captured by $\gtrsim10$ cells and simulations
in the literature do not show strong variations in filament
profiles with resolution \citep[e.g,][]{federrath2016a}.
We also note that the inertial
range of $N=512^3-1024^3$ simulations of $M\approx5$ turbulence is known
to be limited \citep[e.g.,][]{kritsuk2007a,federrath2010a}.
Increasing the numerical resolution of the simulations can 
obviate or mitigate these limitations.

\subsection{Outstanding Issues}

In this work, we have used supersonic isothermal turbulence
to generate a set of dense regions whose properties we have
measured and modeled physically. The methodology for generating the
turbulence and using it as a model for molecular clouds is
well-established, and we follow the approach of previous
works discussed in Section \ref{section:intro}. Once
self-gravity begins to operate, we expect the subsequent
collapse to follow the picture laid out by \citet{murray2015a}
where the adiabatic heating mechanism we identified in 
\citet{robertson2012a} changes the nature of the turbulence
in the clouds as they condense. Adiabatic heating
moderates the collapse, and gives rise to the density and
velocity structure measured in simulations that include
self-gravity and follow the collapse of individual
regions to small scales
\citep[e.g.,][]{kritsuk2011a,murray2017a,mocz2017a,li2017a}.
Once star formation commences, the input of feedback in
various forms \citep{matzner2000a,maclow2004a,krumholz2007a,federrath2015a,raskutti2016a,rosen2016a,li2017a} may influence the overall star formation
efficiency by dispersing low density material.
Although the 
collapse to high densities occurs on short (e.g., several free-fall)
time scales, the cloud itself could
persist much longer if its material is replenished. 
An important
issue then is to determine how the persistent turbulent flow
in star-forming molecular clouds originates or is organized
on large scales.

A possible avenue is the density enhancement of gas passing
through potential perturbations as it orbits in the galactic
disk, such as in spiral arms \citep{shu1973a}, that leads to
a convergent flow \citep[e.g.,][]{hartmann2001a}. 
Various
mechanisms have been envisioned for using the response of the
gas in spiral features to promote the formation of molecular
clouds, such as agglomeration, cooling and
thermal instability, large-scale
Jean instability, and cloud-cloud collisions
\citep{kim2008a,dobbs2008a,dobbs2008b,dobbs2008c,tasker2009a,dobbs2015a},
and to drive turbulence \citep[e.g.,][]{kim2006a}. 
Using the response of gas to galactic potential
perturbations to understand the initial conditions for
persistent turbulent flows of molecular clouds may be promising
if it can explain how the size-line width relation of
originates on large scales and determine what sets the
maximum size of clouds. Once molecular clouds are organized
from more diffuse galactic disk material and placed on the
size - line width relation on the largest scales, the
properties of
the resulting 
turbulent flow, adiabatic heating in interior regions
undergoing gravitational
collapse, and possible input from feedback
may combine to provide a complete picture for star formation
on large scales. We leave analysis of this speculation for
future work.

%
%
\section{Conclusions}
\label{section:conclusions}

Astrophysical fluids often display turbulent motions,
owing to the large Reynolds numbers and low viscosities
typical of conditions observed in molecular clouds, neutral
hydrogen gas, and even the ionized interstellar medium.
The properties of supersonic isothermal turbulence 
especially influence dense molecular clouds with large radiative
cooling efficiencies and bulk motions with velocities in excess of 
the thermal value. As a result, the statistical properties of
supersonic turbulence have provided the core of analytical theories
of the formation rate of populations of stars. Owing in part 
to the complexity of turbulence, these models have largely
emphasized the
statistical properties of the fluid rather than the character of
individual dense regions that may collapse to form stars.

Our work adopts a new approach that aims to describe the properties
of the dense regions in supersonic isothermal turbulence that
serve as the sites of star formation. Our model is grounded by
a distinguishing feature of
supersonic isothermal turbulence, that the origin of large
density contrasts in the turbulent fluid is from
strong shocks with Mach number $\Mach$ 
where the density in post-shock regions
is enhanced by a factor $\sim \Mach^{2}$ relative
to the pre-shock material. The volume of supersonic isothermal 
turbulence occupied by dense fluid is small, and as a result 
after their formation these shocked regions
rapidly encounter lower density
material. The low density external medium applies a 
ram pressure to the shocked regions,
and the thinness of the dense shocked regions
implies that the internal structure of the isothermal shocked regions
will quickly
adjust such that its pressure (and density) gradient balances
the force applied by the ram pressure. Using this local force
balance at the front of the shocked region,
we model the structure of the
fluid behind the shock
as an exponential atmosphere whose rapidly
declining density reflects the pressure gradient necessary to
counter the force from the ram pressure applied across the face of
the shocked region.

We use simulations of supersonic isothermal turbulence to compare
our analytical model with detailed hydrodynamical calculations
of the fluid structure.
We study the density structure of shocked regions,
and find consistency
between their density distributions and our exponential atmosphere
model.
We then use idealized simulations to study the evolution of exponential 
shocked regions
traveling through a low-density background, and find that we can
describe the time-dependent structure of the exponential shocked regions
in terms
of the integrated deceleration of each region
by the ram pressure force
applied by the low density material, the corresponding hydrostatic balance
of this force and the pressure gradient behind the shock,
and
the mass of the background fluid accreted into the shocked region.

After determining that the exponential atmosphere model appears consistent
with the structures of isothermal shocked regions
in our idealized and supersonic
turbulence simulations, we use the results of the model to infer some
additional extensions. Using the catalog of shocked regions
constructed from our
tracer particle distributions, we measure the correlation function of
dense regions and find that it scales as a power-law $\xi(r)\propto r^{-1.5}$
over scales between a few times the grid size to the driving scale. Dense
regions become anti-correlated at about the driving scale, possibly owing to the
large scale rarefactions in the fluid on these scales.
Computing the volumes associated with each dense region from cell tessellations
about the tracer particle locations allows us to demonstrate that the
dense tail of the cumulative density function can be entirely accounted
for through the sum of individual shocked region
structures.

We compute the gravitational potential of the turbulent fluid, and determine
the characteristic density and size of bound regions for a turbulent cloud marginally bound
on a scale comparable to the box size. The global minimum of the gravitational
potential of the turbulent fluid corresponds to a region of moderate density, with
a volume large enough to contain substantial mass (a few tens of percent of the
total cloud). The densest regions in the turbulence correspond to local (not global) potential
minima but do not contain substantial mass as distinct regions.
Without the influence of gravity the larger overdensities will dissipate on their
local Mach crossing time, which we find to be of order $\tcol\sim \Mbar^{-1}L/\cs$.

As this model for the properties of individual shocked regions
in supersonic
turbulence is developed, extended, and modified, the 
time-dependent evolution of the structure of dense regions in turbulence
can be utilized to describe the observed properties of molecular clouds.
Descriptions of the
statistical properties of turbulence and the structural properties of
individual dense regions can be employed together to improve our
physical picture of star formation in supersonically turbulent gas.

\acknowledgments

We thank the anonymous referee for helpful suggestions that improved our manuscript.
Some of these calculations made use of the {\it Hyades} supercomputer
at UCSC, supported by grant NSF AST-1229745. This work used the Extreme Science and Engineering Discovery Environment (XSEDE), which is supported by National Science Foundation grant number ACI-1548562; see \citet{towns2014a}
for more details.

\software{ {\it Athena} \citep{stone2008a}, {\it FLASH} \citep{fryxell2000a},
and {\it Voro++} \citep{rycroft2009a}.}

\appendix

%
%
\section{Tracer Particle Scheme}
\label{section:tracers}

To assist in studying the formation and evolution of shocked regions
in 
our turbulence simulations, we have implemented a
new tracer particle scheme into {\it Athena}.
We begin by placing a single tracer
particle in the center of each cell at the start of the simulation.  The velocity and position of
the particles are updated using a second-order Verlet-like scheme, and during the simulation
the properties of the fluid
at the tracer particle locations are interpolated from the grid using a triangular-shaped cloud (TSC)
scheme \citep[e.g.,][]{hockney1988a}. The tracer particle data are written along with the
grid data for each simulation snapshot. 

Over each time step $\Delta t$,
the tracer particle positions $x$ are evolved according to
\begin{equation}
x_{\mathrm{t}+\Delta\mathrm{t}} = x_{\mathrm{t}} + \frac{1}{2}\left(\vel_{\mathrm{t}+\Delta\mathrm{t}} + \vel_{\mathrm{t}}\right)\Delta\mathrm{t},
\end{equation}
\noindent
where the particle velocities $\vel$ are computed from the 
interpolated fluid mass and momentum densities at times $t$
and $t+\Delta t$.
The Courant condition that determines the
time step criterion in {\it Athena} is also altered
to account for the tracer particle motions.
When calculating the time step from the Courant
condition, the particle velocities are treated analogously
to the signal speed calculated from the fluid
properties on the grid. Since the particle
properties are interpolated from the grid, the
alteration to the Courant condition typically
only has a minor effect on time steps.

The tracer particle scheme is parallelized in
a manner that reflects the communication structure of
{\it Athena} as a whole. Each local grid, typically
evolved by its own parallel process, is assigned
a list of local tracer particles.  As particles
cross grid interfaces, the grid boundary conditions
are applied.  If the tracers cross an internal boundary
between grids owned by different {\tt MPI} processes
or periodic boundaries on the exterior of the computational
volume, they are marked, packed
into communication buffers, and then
exchanged with adjacent grids via non-blocking
{\tt MPI\_Isend()}. Grids receive tracers
flowing from adjacent grids on a first-come-first-served
basis using {\tt MPI\_Waitany()} and {\tt MPI\_Irecv()}
calls.  The process is repeated to allow for tracers
to cross multiple cell faces (a rare occurrence very
close to grid edges and corners). If necessary,
the tracer positions
are then adjusted appropriately for wrapping across
periodic boundaries.

%
%
\section{Interpolation of Tracer Properties}
\label{section:reinterpolation}

Godunov-style grid-based fluid simulations use reconstruction
methods to provide subcell gradients in the
fluid, and thereby improve extrapolations to the cell
interfaces where the Riemann problems are computed
for flux calculations. The values of the fluid on 
the grid are often stored as central cell-averages,
and correspond to integrals over the fluid shape within
the cell in a finite-volume approximation. Given
the irregular distribution of the tracer particles,
interpolation methods are used to estimate the
fluid properties at the tracer locations.
In computing the simulations presented here, 
to evolve the tracer particle positions and
velocities
the fluid properties simulated
on the grid are interpolated at the tracer particles
using the TSC method. 
Since this interpolation
method acts like a kernel-weighted estimate of the
fluid properties at locations intermediate between the
grid locations, the fluid properties are smoothed
on scales comparable to the grid spacing. A consequence
of this interpolation scheme is that the tracer particle
properties near extrema in the fluid are smoothly truncated.
If the particles are to be used to reconstruct the statistical
properties of the fluid simulated on the grid, such as the
density probability density function (PDF) measured
volumetrically, then the extrema
must be preserved. To maintain these features of the fluid
properties interpolated at the tracer particle locations,
we require different interpolation schemes than TSC.
A trivial choice
would be to use a nearest grid point method, but this 
method reduces the order of the interpolation and does
not reflect the assumed model of the original grid 
simulation that subcell gradients in the fluid properties exist.
Below, we introduce two additional interpolation methods that 
work well and are comparable in quality, and better capture the extrema
in the simulated fluid properties. 

In re-interpolating the
fluid at the tracer particle positions, we treat the
tracers as sample locations for constructing new approximations
to the smooth fluid properties between grid cells. We
do not treat the tracers as Lagrangian elements \citep[see, e.g.,][]{genel2013a},
and do not assign unique mass elements to tracers in an attempt to
follow parcels of gas.
However, since we are concerned primarily with the 
development and evolution of shocked regions
(which are not
Lagrangian structures) this issue poses no serious
difficulty. When necessary and discussed in 
more detail below, shocked regions
are followed using
successive generations of tracer particles that move
through the structures of interest. Shock
fronts
can be identified based on the time-dependent density
of the fluid as sampled by the tracers.

%
%
\subsection{Piecewise Parabolic Interpolation}
\label{section:reinterpolation:ppi}

Spatial reconstruction using piecewise interpolation
is a core feature of modern Godunov-based
fluid dynamics methods. The piecewise
parabolic method \citep[PPM;][]{colella1984a}
is one such scheme that provides a third-order
accurate reconstruction on a one-dimensional
grid by using cell averages across a stencil
to constrain the assumed 
parabolic shape of the fluid properties
interior to each cell. PPM is often
used in grid-based simulations
when extrapolating the fluid properties
at the cell interfaces to compute the
input states for a Riemann problem during
the hydrodynamical time integration.
This treatment is approximate, and the
reconstruction in each direction is usually
treated independently, such that the 
PPM reconstruction is determined from
a grid-aligned stencil of cells in a 
single row or column. However, suitable
averages of one-dimensional PPM reconstructions
in each direction can be taken to provide
an approximate
three-dimensional interpolation scheme
that we call piecewise parabolic interpolation.

Consider a one-dimensional PPM reconstruction
$q_i(x_i)$ providing the assume parabolic shape of
the fluid at location $x_i$ within a cell along
direction $i\in [x,y,z]$. For $q_i$, we
adopt the same PPM reconstruction used by the
FLASH code and refer the reader to \citet{fryxell2000a}
for details.  To construct an estimated three-dimensional
interpolation $q(x,y,z)$ within a cell centered
at $[0,0,0]$, we then compute
a simple linear interpolation
between the $x-$, $y-$, and $z-$interpolations 
using the heuristic
\begin{eqnarray}
\label{eqn:ppi}
\phi &=& \arctan\left(|x|,|y|\right)\nonumber\\
\theta &=& \arctan\left(|z|,\sqrt{x^2 + y^2}\right)\nonumber\\
q_{xy} &=& q_x \left(1-\frac{2\phi}{\pi}\right) + q_y \frac{2\phi}{\pi} \nonumber\\
q(x,y,z) &=& q_{xy}\left(1-\frac{2\theta}{\pi}\right) + q_z \frac{2\theta}{\pi}.
\end{eqnarray}
\noindent
We refer to this as Piecewise Parabolic Interpolation (PPI) in the text.
We use the same slope limiters and
monotonicity constraints as \citet{fryxell2000a},
which result in a
constant reconstruction within cells at local
extrema. Interpolations along the $x-$, $y-$
or $z-$directions relative to the cell centers
reduce to the corresponding
one-dimensional PPM reconstructions. Other
three-dimensional averages of the one-dimensional
PPM reconstructions are possible and may prove
more advantageous for other applications.

%
%
\subsection{Gaussian Process Interpolation}
\label{section:reinterpolation:gpi}

When interpolating the fluid properties from the 
grid, using a method constrained to reproduce the
fluid properties at the grid centers can help
reproduce extrema. Such a method would
interpolate between
the cell-averaged quantities of the original
fluid reconstruction, but may not reproduce those
quantities when averaged over the grid. If the
interpolation is smooth then such a method can
overshoot extrema on the original grid over a
portion of a grid cell, but appropriate choices
can make such overshoots small in practice.

A smooth interpolation scheme that can be
constrained to pass through central cell
values is Gaussian process interpolation \citep{rasmussen2006a}.
If we treat the cell values in a simulation as
realizations from a Gaussian process,
then we can interpolate between cell values by 
defining a model for the covariance between nearby cell 
values and computing the mean of the Gaussian process everywhere
\citep[e.g.,][]{reyes2016a}.
The method is intrinsically multidimensional, and 
does not require information about the directionality
of the interpolation.

Consider $f(x)$
to be the realization of a function at the cell center locations 
$x$. To interpolate between $x$ and some other location $x_\star$,
one can treat the function as a Gaussian process and 
use the estimate
\begin{equation}
\label{eqn:gpi}
\langle f(x_\star)\rangle\simeq k(x_\star,x)k^{-1}(x,x)f(x)
\end{equation}
where $k(x_a,x_b)$
is a model for the covariance matrix between
the function values at locations $x_a$ and $x_b$,
and $k^{-1}$ is its inverse. Note $k^{-1}(x,x)$
is computed from the covariance of the function as sampled
at grid locations $x$, and has a size $n_x \times n_x$
where $n_x$ is the number of contributing nearby grid cells. 
The interpolants $\langle f(x_\star)\rangle$
are therefore weighted sums of the function values at the
cell centers, but the product of the covariance matrix
with its inverse ensures that $\langle f(x_\star)\rangle\to f(x)$ as
$x_\star\to x$.

The implementation of this scheme involves choices for the
set of cell centers contribute to the interpolant at each
location $x_\star$ and the form of the covariance
function. 
A common model for the covariance function is a Gaussian
\begin{equation}
k(x_a,x_b)=\frac{1}{\sqrt{2\pi l^2}} \exp\left[ - \frac{(x_a-x_b)^2}{2l^2} \right]
\end{equation}
where $l$ is a typical correlation length and $|x_a-x_b|$
is the vector magnitude between the (three-dimensional) locations $x_a$ and $x_b$.
Nominally, given this covariance function all cells in the simulation
would contribute to the Gaussian process interpolation at any interior location.
Since the contributions of cells at a distance more than the correlation length are exponentially
suppressed, only relatively nearby cells within a few correlation lengths need
to be considered.
In practice we choose a correlation length $l=\sqrt{3}\Delta x$, where $\Delta x$
is the cell width, and consider contributions from cells at distances $|x-x_\star|\lesssim 4l$.
Our interpolation stencil is therefore $9^3 = 729$ cells, and our inverse correlation
matrix has $729\times729 = 531441$ elements. However, the regular spacing of the grid
allows for $k^{-1}(x,x)$ to be computed only once and reused for each interpolation.
We use the Intel Math Kernel Library implementation of the Cholesky decomposition to
perform efficiently the matrix inversion. Equation \ref{eqn:gpi} is then evaluated for all $N=512^3$ or $N=1024^3$
tracers at every snapshot of interest.

%
%
\subsection{Comparison of Interpolation Methods}
\label{section:reinterpolation:comparison}

\begin{figure*}[ht]
\begin{center}
\includegraphics[width=7.1in]{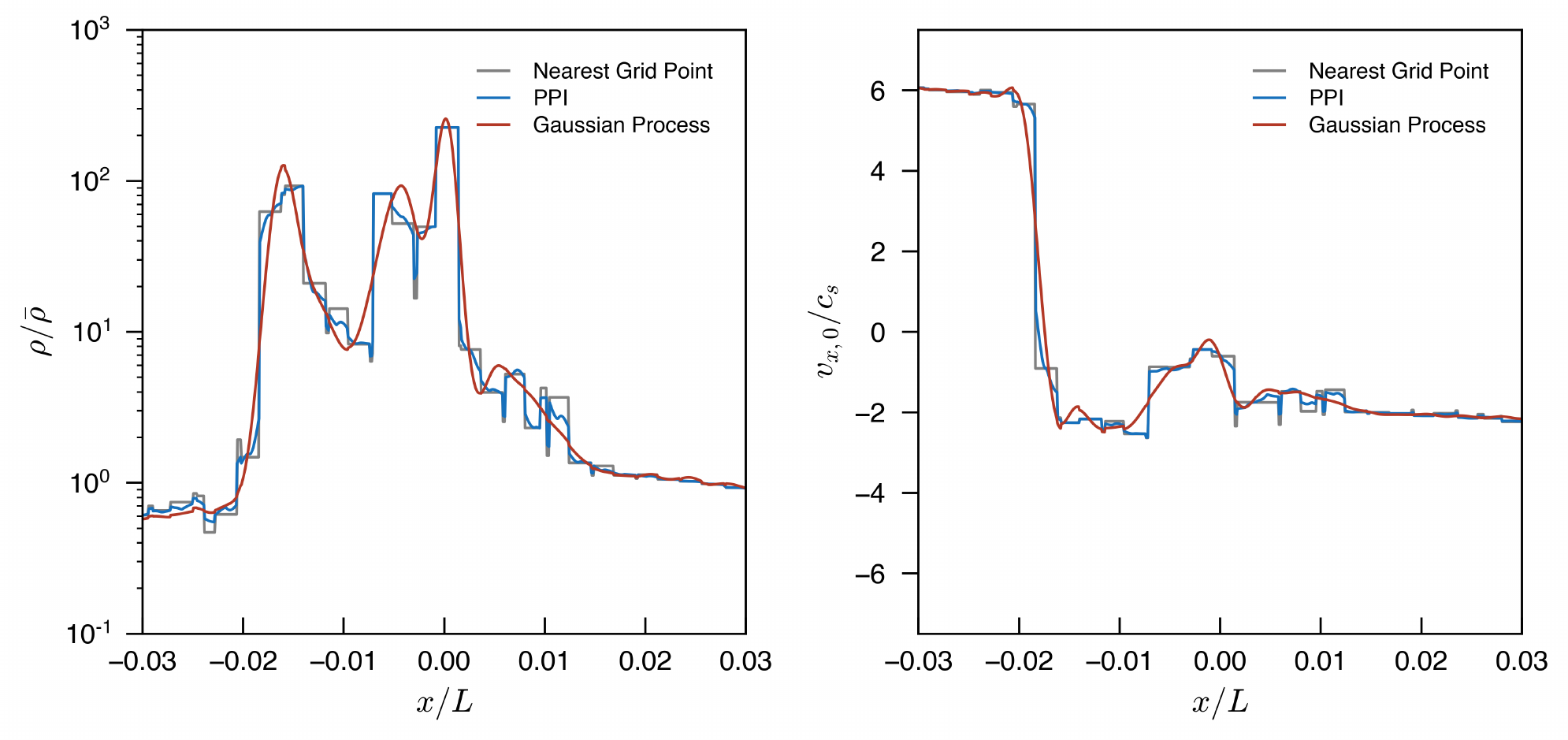}
\caption{\label{fig:interpolation_comparison}
Comparison of interpolations methods. Shown are the Nearest Grid Point (NGP; gray),
Piecewise Parabolic (PPI; blue line), and Gaussian Process (GPI; red line) interpolations of the
density (left) and $x_0$-direction (grid-aligned) velocity (right) field near a dense shocked region
with $\rho\approx230\rhobar$ in the $N=512^3$ simulation. The trajectory of 
the interpolation is not aligned with the simulation grid, such that the
Nearest Grid Point interpolation can show large variations on scales smaller than
the grid scale $\Delta x = L/512$. The PPI method matches the NGP method near the density maximum , but varies more smoothly than NGP. The GPI method is smoother than PPM, as it kernel-weights information from a $9^3$ stencil about each position, but provides a density estimate near extrema within about $10\%$ of the
NGP value.
}
\end{center}
\end{figure*}

The PPI (Section \ref{section:reinterpolation:ppi}) and GPI (Section \ref{section:reinterpolation:gpi}) methods for estimating the simulated fluid properties
reconstructed at the positions of tracer particles provide two approaches with complementary
advantages. Figure \ref{fig:interpolation_comparison} compares the reconstructions provided
by the two methods (PPI in blue, GPI in red), along with the Nearest Grid Point method (gray), for density (left) and velocity (right) interpolation along a skewer
oriented perpendicular to a dense shocked region
identified in the turbulence simulation.
The PPI method preserves the behavior of NGP interpolation in that
density maxima are flattened to a constant within a cell, which allows for the recovery
of the densest part of the density PDF (see Section \ref{section:dense_pdf}). The GPI method
The GPI method provides a smoother interpolation, which is advantageous in determining which
tracer particles to assign to a dense region (Section \ref{section:group_finding}) and in
measuring density profiles behind shocks
(Section \ref{section:density_profiles}), but still recovers the value of the
density maximum to within
$\approx10\%$. We note that the small-scale variations on scales less than the grid
scale $\Delta x = L/512$ in the NGP and PPI methods
arise from the oblique of the interpolation path relative to the grid.

%
%
\section{Group Finding Algorithm}
\label{section:group_finding}

Calculating the properties of individual dense regions
in an Eulerian simulation is straightforward at
any specific time.  For an isothermal fluid,
shock
fonts may be simply identified by their large
density gradient upwind from a large density
enhancement. Regions with a large density enhancement
(e.g., $\rho/\rhobar > M$) must be generated by shocks.
However, if one wants to monitor the formation
and evolution of dense regions an Eulerian scheme presents
some difficulties. When tracking the history of an
individual shocked
region 
back in time, the associated shock
will inevitably have originated
from the compression of two distinct regions.  The
history of the shocked region
before this point will become
indeterminate.  Very dense shocked regions
in isothermal
turbulence are themselves created from the interaction of
lower density shocked regions,
and following the formation of
such higher generation shocked regions
quickly becomes 
difficult in an Eulerian simulation.

By using tracer particles, we can monitor the motion of
particles (at least momentarily) associated with a 
given shock.
Since the tracers move through the
shocks,
we can use successive generations of tracers
to follow the motion of a shocked region
with time and reconstruct
its history (see Section \ref{section:group_tracking} below).
To achieve this, we must have a robust method for
associating groups of tracer particles in shocked regions.
There are a variety of possible methods and we have tried
several, but we find the
following algorithm to work well in practice.

Since we are interested in shocked regions,
we first select
particles denser than a chosen threshold $\rhothresh$
to incorporate into distinct regions.  This simple
selection essentially defines the particle membership
of regions we study, but to follow the formation and
evolution of the shocked regions
we
need to organize the tracers into distinct groups.
Our method for organizing the dense tracers into individual
shocked regions
is a friends-of-friends (FOF) group finding algorithm
(e.g., \citealt{huchra1982a}), where we separate
tracer particles into FOF groups based on their
spatial proximity. Regions whose closest tracers are
separated by more than a linking length $b$ are considered to
be distinct.  We choose the linking length to be of order
the expected mean inter-particle separation for a density
of $\rho \sim \rhothresh$.
By conditioning the FOF membership on density, we avoid 
using the linking length as an estimator for the
density near the bounding surface of the shock.
Instead, the
linking length is simply used as a metric to determine whether
two shocked
regions are proximate or distant.
This feature of our algorithm
avoids some of the FOF group finding pathologies 
discussed in \citet{more2011a} in the context of cosmological
N-body simulations.

Our algorithm for identifying density-conditioned FOF groups is
as follows:
\begin{enumerate}

\item {\bf Select Tracers Above a Given Density Threshold}. We select tracers
above a density threshold $\rhothresh$ to define the
spatial extent of shocked regions.
The tracer densities
are computed by interpolating the local fluid properties, and
placing a condition on the tracer densities for FOF group
membership amounts to selecting regions within isodensity
contours sampled at the tracer locations. Conditioning the
group membership on density also reduces the
number of tracers that need to be processed, which 
reduces the computational time of the rest of the algorithm.

\item {\bf Construct a Graph}. The separation of 
dense tracers into distinct regions requires an evaluation
of their spatial distribution.  More specifically, we
require a method for determining whether tracers are
proximate or well-separated. By constructing a graph of
the tracers,
the distances between nearby tracers are recorded
and can be compared with a metric to group tracers
based on their proximity. 
If we consider the tracer positions as vertex locations in
the graph, we generically do not know which 
edge connections between vertexes with lengths $l<b$
will comprise connections between members of a shocked region.
Constructing a complete graph of the tracer distribution would
guarantee that all desired connections would be identified, but
constructing such a graph for $N$ tracers
requires $O(N^2)$ operations and is
computationally prohibitive. Instead, we construct a graph of
the tracers where each tracer (vertex) is connected by some
number $\NST\ll N$ edges to other vertices.  We require these
edges to be among the shortest available connections, since
all edges with lengths $l>b$ will eventually be discarded
once the FOF groups are constructed. A nearest neighbor search
is therefore performed, with the shortest $\NST$ edges
for each vertex identified. To construct this graph, a kD-tree
of the tracer positions is generated and utilized to perform
rapidly the neighbor searching. A value of $\NST\sim10$ is chosen to
provide a large connectivity between regions (we do not require
that all tracers are connected in the graph, we just require that
all tracers closer than a distance $b$ are connected).  Picking
too small a value of $\NST$ can result in disconnected regions
whose closest $\NST$ neighbors are self-contained, while picking 
too large a value of $\NST$ increases the computational cost of the
algorithm. Some of the connections may have a length $l>b$, and will
be removed later.  
This step of the algorithm is illustrated in the
left column of Figure \ref{fig:spanning_tree}.

\begin{figure*}[t]
\begin{center}
\includegraphics[width=7.1in]{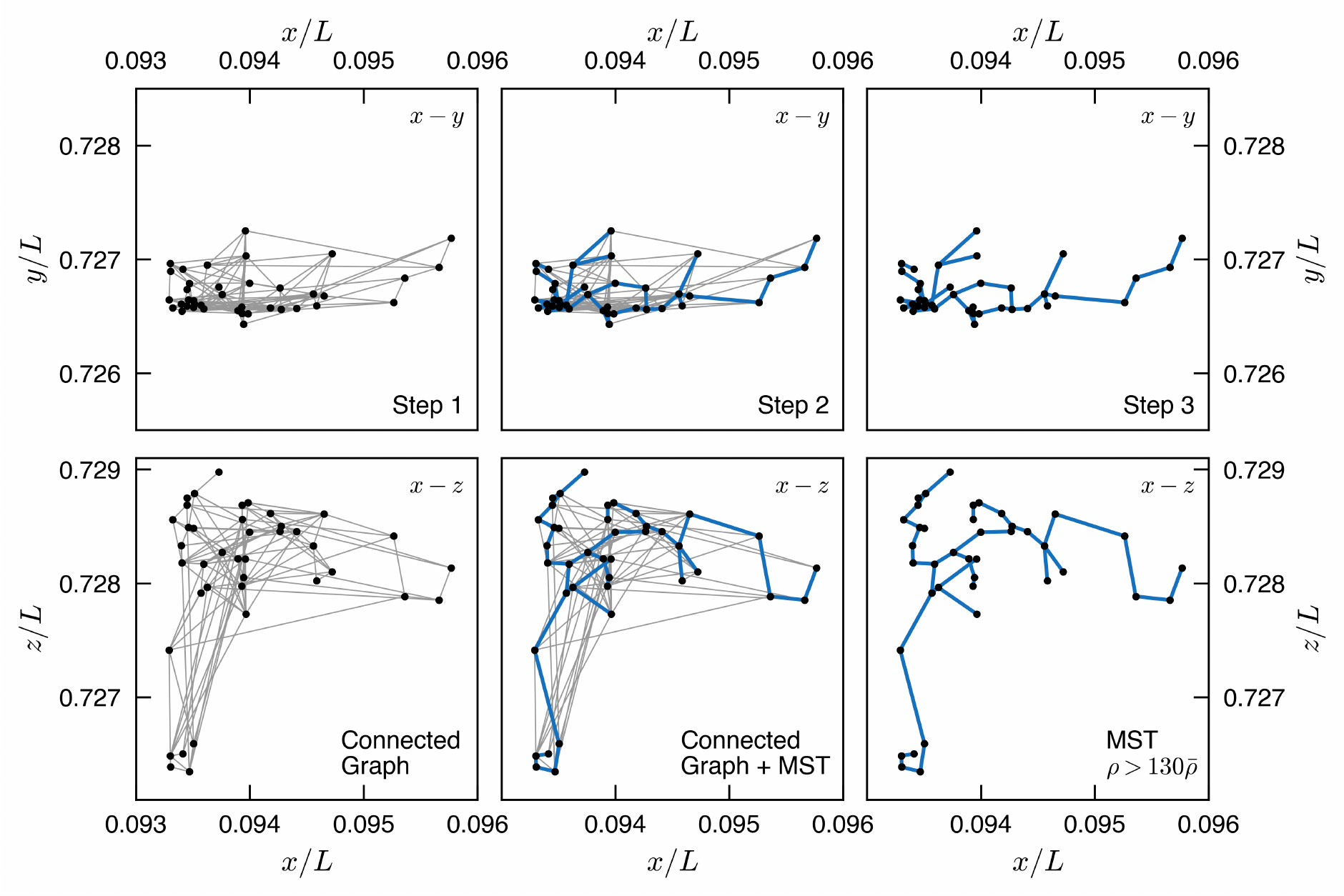}
\caption{\label{fig:spanning_tree}
Algorithm for rapid construction of FOF groups with linking length $b$
from minimum spanning trees.  A kD-tree 
is constructed based on the particle positions, and for each particle
the kD-tree is used to search for the nearest $\NST$ particles.  The
edges formed by the connections between neighbors are added to a graph.
Locally, for each group, these 
edges form a graph
(left column) comprised 
of the $\NST$ shortest connections between each vertex and
its neighbors. The $x$-$y$ (upper row) and $x$-$z$ (lower row) projections
are shown for each graph.  
The minimum spanning tree (MST) of each graph's particles is then
constructed (blue lines, middle column) using 
the algorithm of \citet{kruskal1956a}.  The remaining
edges with lengths $l>b$ are then discarded (right column), leaving behind the MST.
As a final step, since the algorithm operates on multiple subvolumes of the simulation in 
parallel MSTs spanning subvolume boundaries are stitched together.  
The end result is a set of friends-of-friends groups with linking length $b$,
each organized into a MST.
}
\end{center}
\end{figure*}

\item {\bf Construct Minimum Spanning Trees}. The
use of minimum spanning trees (MSTs) to define FOF groups
is a standard technique \citep[see, e.g.,][]{knebe2011a}.
With a sufficiently large choice for $\NST$ in the graph construction
step, the graph of each dense region will contain the MST of
its vertices with edges of lengths $l<b$.  The edges comprising the
MST are identified using the algorithm of \citet{kruskal1956a}.
Briefly, starting with the graph of a connected region, edges are
added to the tree from shortest to longest. At each insertion,
the shortest available edge that does not form a loop in the tree
is added. Determining whether adding an edge to the tree would result
in a loop amounts to knowing whether the vertices at the ends of the
edge already belong to the tree. This issue is complicated somewhat by the
\citet{kruskal1956a} algorithm, which effectively involves the 
merging of subtrees. When constructing the MST, we therefore create a linked
list of vertices that belong to a subtree that is ``owned'' by its densest
tracer. Each element in the subtree list contains a pointer to the first element,
which in turn tracks the tracer that owns the subtree.  When deciding whether 
to insert an edge,
the possible membership of both vertices of the edge can be quickly checked
and the insertion vetoed if both vertices are already owned by the same tracer
and belong to the same subtree.  Otherwise, the subtrees can be merged (with the
densest member tracer owning the newly merged tree) or a new subtree created if neither
vertex has yet been inserted.  The net result of the algorithm is to produce
a forest of MSTs that are separated by no less than a distance $b$.
This step of the algorithm is illustrated in the
middle column of Figure \ref{fig:spanning_tree}.

\item {\bf Discard Excess Edges and Retain FOF Groups}. The FOF group defined
by a linking length $b$ can be constructed from the MST comprised of edges
with lengths $l<b$. The algorithm to this point constructs a forest of MSTs
whose longest lengths may be $l>b$, so each group is examined and split
appropriately by removing any edges that are too long. 
This step of the algorithm is illustrated in the
right column of Figure \ref{fig:spanning_tree}.
Again, it should be
emphasized that by conditioning the FOF group membership of the tracers on
density, distinct regions whose isodensity contours for $\rho\ge\rhothresh$
do not intersect will remain as separate FOF groups in this algorithm provided 
the linking length $b$ is not too large.
Figure \ref{fig:mst_with_density} illustrates the relative separation of
an example set of FOF groups identified by the algorithm.  In this case, the
linking length would have to be increased dramatically for these regions to
be merged into a single FOF group.  The algorithm is therefore expected to be
robust to order unity changes in $b$, which we have confirmed in tests.

\item {\bf Stitch Subvolumes}. The spatial extent of shocked
regions varies
considerably based on the threshold density $\rhothresh$, and for a simulation
comprised of subvolumes evolved by parallel processes, connected regions may
frequently span the boundaries between computational subvolumes. Since the
FOF groups are initially constructed for each computational subvolume in parallel,
some groups may need to be ``stitched'' together.  For simplicity, this stitching
step is performed serially.

\item {\bf Repeat with Different Density Thresholds}. The
group finding algorithm is repeated for a hierarchy of density thresholds,
allowing for density enhancements within lower density groups to be
identified as distinct local density maxima. If multiple distinct density
peaks are identified within a region, lower density particles are assigned
to the peak containing the most proximate tracer particle of higher density.
\end{enumerate}

Figure \ref{fig:mst_with_density} provides an illustration of the above algorithm.
The complex density field is apparent in the top panel that shows a logarithmic
projection through one-fourth of the volume. The tracer particles moving in 
response to the fluid properties become concentrated in dense regions (lower
right panel). Given an threshold density, in this case $\rho\ge 15\rhobar$, the
algorithm groups tracer particles in dense regions into convenient minimum spanning trees
(lower left panel). Appendix \ref{section:shock_orientation} describes
how these tracer particles groups are used to define interpolation trajectories for
measuring the fluid properties near shocks,
and Appendix \ref{section:group_tracking}
details how the groups are tracked with time
during the simulation.

\begin{figure*}
\begin{center}
\includegraphics[width=7.1in]{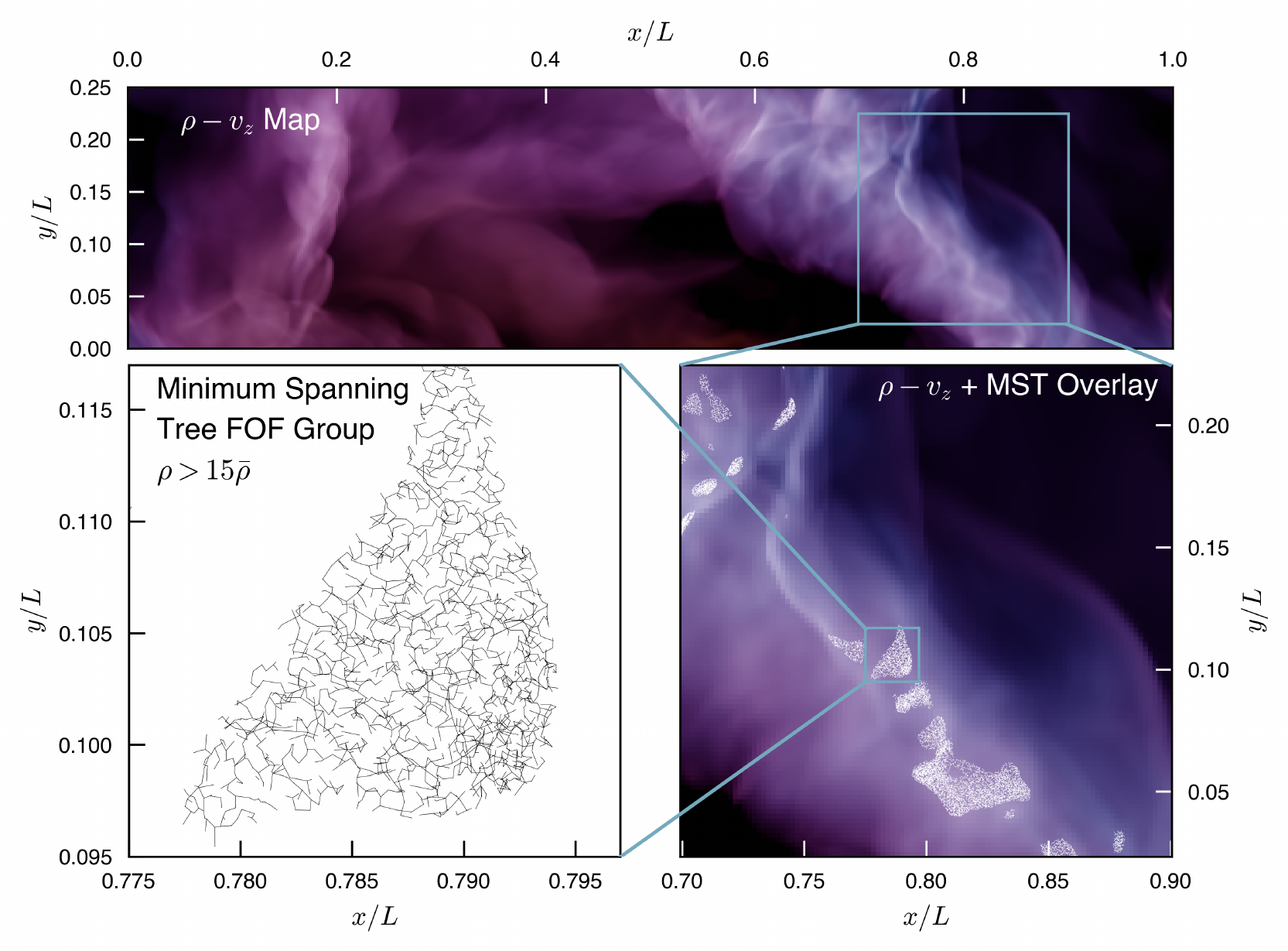}
\caption{\label{fig:mst_with_density}
Spatial relation between Friends-of-Friends (FOF) groups and overdense regions in 
the supersonically turbulent fluid.  The fluid is very inhomogeneous, with
large variations in the density.  The upper panel shows a logarithmic projection
of the density field (intensity) through one fourth of the simulation volume, 
color coded by the $y$-velocity of the fluid 
(positive velocities are red, negative velocities are blue).  Zooming in on
a region with large projected density (lower right panel), the filamentary appearance
caused by the projection of dense sheets is apparent.  The minimum spanning trees
comprising the FOF groups (white, constructed for particles with densities $\rho/\rhobar \ge 15$)
is overlaid to show regions of large three dimensional density.  Zooming in further
on a handful of FOF groups (bottom left panel), the distinct character of the dense
regions can be seen.  The relative separation between groups is considerably larger than
their interior mean inter-particle separation.
}
\end{center}
\end{figure*}

%
%
\section{Shock Orientation and Trajectory Interpolation}
\label{section:shock_orientation}

By selecting tracer particles with densities above some chosen threshold, dense regions 
in the simulations can be identified and their properties measured.  Since regions with
densities $\rho\gg\rhobar$ in isothermal turbulence must be generated by shocks,
we are interested in measuring the properties around the dense regions in turbulence.
Once the dense regions are identified from the tracer particles selected above a
given density threshold, their structural properties are estimated from the density
distribution probed by the tracers.

To measure a density and relative velocity profile about each shock,
an estimate of
the local moment of inertia tensor is constructed.
For each shocked region,
the densest
tracer particle is identified.  The location of the shocked region peak
is taken to be the
average position of particles in the tenth percentile of density that are in the
closest quintile of particles in separation from the densest tracer.
A density-weighted estimate of the moment of inertia tensor is constructed from these $n_p$ tracer
particles as

\begin{equation}
\label{eqn:inertia}
\mathbf{I} = \frac{1}{\Sigma_{\rho}}\left[ 
\begin{array}{ccc}
\sum \rho_{i} \left(y_{i}^{2} + z_{i}^{2}\right) & - \sum \rho_{i} x_{i} y_{i} & -\sum \rho_{i} x_{i} z_{i} \\
-\sum \rho_{i} y_{i} x_{i}                       & \sum \rho_{i} \left(x_{i}^{2} + z_{i}^{2}\right) & -\sum \rho_{i} y_{i} z_{i} \\
-\sum \rho_{i} z_{i} x_{i} & -\sum \rho_{i} z_{i} y_{i}  & \sum \rho_{i} \left(x_{i}^{2} + y_{i}^{2}\right)
\end{array}
\right] 
\end{equation}
\noindent
where $\rho_i$, $x_i$, $y_i$, and $z_i$ are the density and $x$, $y$, and $z$ positions of the 
tracer particles, the summations span the range $i\in[0,n_p)$ and we abbreviate the sum of the 
densities as
\begin{equation}
\Sigma_{\rho} = \sum_{i=0}^{n_p-1} \rho_{i}.
\end{equation}
\noindent
The estimated moment of inertia tensor is then diagonalized to find the principal axes of the
density distribution near each shocked region peak.
We will use $\vxi$ to denote the
eigenvector corresponding to the largest eigenvalue of the moment of inertia tensor.  If the
dense regions are flattened into sheet-like shocked regions,
then we would expect $\vxi$ to be oriented
perpendicular to the shock face.
Since the eigenvectors are not uniquely defined to within a
reflection, we select the orientation of $\vxi$ that has a positive projection along the bulk
velocity of the shocked region.
When tracking the shocked region
with time (see below), we choose a consistent
orientation of the principal axes.

Once $\vxi$ is determined, we can parameterize a path through the fluid simulated on the
Eulerian grid that runs through the shock
parallel to $\vxi$.  In terms of some path length
$\eta$, the equation of this line is
\begin{eqnarray}
\vec{x} = x_{0} + \eta\left(\vxi\cdot\hat{i}\right)\hat{i} \\
\vec{y} = y_{0} + \eta\left(\vxi\cdot\hat{j}\right)\hat{j} \\
\vec{z} = z_{0} + \eta\left(\vxi\cdot\hat{k}\right)\hat{k},
\end{eqnarray}
\noindent
where $x_{0}$, $y_0$, and $z_0$ correspond to the three-dimensional location of the peak about which 
the moment of inertia tensor is defined.
The properties of the fluid along this path near each dense region can be interpolated from the grid
for each position $[\vec{x}(\eta),\vec{y}(\eta) ,\vec{z}(\eta)]$.

%
%
\section{Shock Tracking Algorithm}
\label{section:group_tracking}

Given the methodology for defining shocked
regions described in Appendix \ref{section:group_finding},
we can engineer a fast parallel scheme for
reconstructing the history of shocked regions
from a suite of catalogues
generated for
a sequence of times during the simulation.  Assuming
the computational domain is divided into adjacent
subregions assigned to separate parallel processes, we
have found the following simple divide-and-conquer 
algorithm to work well.  Other schemes based on
identifying velocity jumps have also been shown to
be effective \citep[e.g.,][]{smith2000a,smith2000b}.

In our approach, each parallel process compares
the shocked regions
identified at time $t_1$ with
shocked regions
in the same and adjacent regions at an earlier
time $t_0$, examining only shocked regions
that could have
shared particles over the elapsed time $t_1-t_0$
given their relative velocities. Regions that contain
the same particles at different times are incorporated
into a recorded history for each region.

Depending on the shock
finding algorithm, analogues of
the common pathologies seen in dark matter halo
merger tree reconstructions \citep[see, e.g.,][]{fakhouri2008a} are also encountered in
the histories of shocked regions.
For instance, since our approach
is to use a FOF criterion conditioned on particle density
for determining the membership of a tracer particle 
in a given shocked region,
if the topological structure
of a shocked region
includes a narrow ``bridge'' the
FOF finding algorithm can fragment regions into 
seemingly unassociated structures at a later or earlier time if
the density in the bridge varies with time.
Owing to the complicated topology of turbulence, this fragmentation
can be physical in origin as the connectivity of isodensity contours
can vary over short time scales.
As we argued in Appendix \ref{section:group_finding} the algorithm is
robust to choices in the linking length at any one time, but it is
reasonable to expect that derived quantities like 
the ``merging time'' of two distinct shocked regions
or the possible fragmentation
of groups across two time steps
will depend on the choice of linking length.
Nonetheless, even given these considerations, robust reconstruction of a shocked region's
history can be achieved by either monitoring ``ancestor''
shocked regions
in multiple
snapshots at different times simultaneously or, as we
prefer, by tracking the histories of all ancestor 
shocked regions
at all previous times and reconstructing the
``main branch'' of the history after the fact.

With an additional step, the method can also be used 
to reconstruct the
history of fluid surrounding the shocked region
and its predecessors.  
One option
is to find all tracers
associated with the shocked regions
at any time, and
then read their properties 
at all times in the recorded history from the
tracer particle snapshot files.  A second option is
to use the history of the shocked regions to define
volumes co-moving with the structures of interest
 and reconstruct
the fluid properties within those volumes from either
the tracer particle samples or the Eulerian grid at
each time.  We have used both these methods for various
measurements presented in this work.

\end{document}